\documentclass{vldb}

\usepackage{graphicx}
\usepackage{balance}  
\usepackage{amsmath,amssymb}
\usepackage{xspace}
\usepackage[table,xcdraw]{xcolor}
\usepackage{array,multirow}
\usepackage{diagbox}
\usepackage{setspace}
\usepackage{lscape}
\usepackage{amsthm}
\usepackage{lineno}
\usepackage[normalem]{ulem}
\usepackage{pgfplots}
\pgfplotsset{compat=newest}
\usepackage{algorithm}
\usepackage[noend]{algpseudocode}
\usepackage{hyperref}

\usepackage[scaled=.7]{beramono}
\usepackage[T1]{fontenc}

\pgfplotscreateplotcyclelist{mycolorlist}{%
orange,every mark/.append style={fill=blue!80!black},mark=triangle\\%
red,every mark/.append style={fill=red!80!black},mark=star\\%
brown!60!black,every mark/.append style={fill=brown!80!black},mark=otimes\\%
black,mark=star\\%
blue,every mark/.append style={fill=blue!80!black},mark=diamond*\\%
red,densely dashed,every mark/.append style={solid,fill=red!80!black},mark=*\\%
brown!60!black,densely dashed,every mark/.append style={
solid,fill=brown!80!black},mark=square*\\%
black,densely dashed,every mark/.append style={solid,fill=gray},mark=otimes*\\%
blue,densely dashed,mark=star,every mark/.append style=solid\\%
red,densely dashed,every mark/.append style={solid,fill=red!80!black},mark=diamond*\\%
}

\newcommand{\methodName}{Kensho\xspace}

\newcommand{\stepTwo}{semantic association discovery\xspace}

\newcommand{\finalAssociation}{semantic association\xspace}
\newcommand{\eat}[1]{}
\newcommand{\length}{\mbox{$\cal L$}}
\newcommand{\sa}[1]{\mbox{$\mathcal A (#1)$}}

\newcommand{\basic}{baseline}
\newcommand{\rel}{r2r}
\newcommand{\atr}{c2a}

\theoremstyle{definition}
\newtheorem{definition}{Definition}[section]
\newtheorem{example}[definition]{Example}

\newcommand{\squishlist}{
 \begin{list}{$\bullet$}
  { \setlength{\itemsep}{0pt}
     \setlength{\parsep}{1pt}
     \setlength{\topsep}{1pt}
     \setlength{\partopsep}{0pt}
     \setlength{\leftmargin}{1em}
     \setlength{\labelwidth}{1em}
     \setlength{\labelsep}{0.5em} } }

\newcommand{\squishend}{
  \end{list}
}

\vldbTitle{}
\vldbAuthors{}
\vldbDOI{}
\vldbVolume{}
\vldbNumber{}
\vldbYear{}

\begin{document}
\sloppy

\setlength\linenumbersep{-0.25cm}
\definecolor{darkgreen}{rgb}{0.1,0.6,0.1}
\definecolor{pink}{rgb}{0.858, 0.188, 0.478}
\definecolor{brown}{rgb}{0.59, 0.29, 0.22}
\definecolor{bluegreen}{rgb}{0.1, 0.6, 0.6}

\newcommand{\TRep}[1]{\textcolor{black}{#1}}

\setlength{\topskip}{0pt}
\setlength{\topsep}{0pt}


\toappear{} 


\title{Knowledge Translation: Extended Technical Report}



%
%
%
%

\numberofauthors{4} 

\author{
%
%
\alignauthor
Bahar Ghadiri Bashardoost\\
\affaddr{University of Toronto}\\
\email{ghadiri@cs.toronto.edu}
\alignauthor
Ren{\'e}e J. Miller\\
       \affaddr{Northeastern University}\\
       \email{miller@northeastern.edu}
\alignauthor
Kelly Lyons\\
\affaddr{University of Toronto}\\
\email{klyons@cs.toronto.edu}
\and  
\alignauthor
Fatemeh Nargesian\\
\affaddr{University of Rochester}\\
\email{fnargesian@rochester.edu}}

\makeatletter
\def\@copyrightspace{
    \@float{copyrightbox}[b]
   \begin{center}
        \setlength{\unitlength}{0.2pc}
            \begin{picture}(100,6) 
                \put(0,-0.95){\crnotice{\@toappear}}
          \end{picture}
        \end{center}
    \end@float}
\makeatother

\maketitle

\begin{abstract}
We introduce Kensho, a tool for generating mapping rules between two Knowledge Bases (KBs). To create the mapping rules, Kensho starts with a set of correspondences and enriches them with additional semantic information automatically identified from the structure and constraints of the KBs. Our approach works in two phases. In the first phase, semantic associations between resources of each KB are captured. In the second phase, mapping rules are generated by interpreting the correspondences in a way that respects the discovered semantic associations among elements of each KB. Kensho's mapping rules are expressed using SPARQL queries and can be used directly to exchange knowledge from source to target.
Kensho is able to automatically rank the generated mapping rules using a set of heuristics. We present an experimental evaluation of Kensho and assess our mapping generation and ranking strategies using more than 50 synthesized and real world settings, 
chosen to showcase some of the most important applications of knowledge translation. In addition, we use three existing benchmarks to demonstrate Kensho's ability to deal with different mapping scenarios.
\end{abstract}

\section{Introduction}\label{sec:introduction}
Knowledge bases (KBs) have become building blocks for many knowledge-rich applications. As a result, significant work has been devoted to studying methods for creating and populating KBs. 
Most of this effort is focused on approaches for 
general-purpose KBs and 
less work has considered populating domain-specific KBs. 
Massive numbers of KBs are available today and their numbers are growing making it now possible and desirable to populate (or augment) new KBs with knowledge already available in others.
Currently, the Linked Open Data Cloud (LOD)~\cite{lod} 
contains 1255 KBs each of which contains more than 1000 triples. Together, KBs in the LOD contain billions of triples. 
Many approaches have been developed that facilitate the discovery and recommendation of KBs 
(see~\cite{kolbe2019linked} for a recent survey). While these approaches can help discover a desirable source of knowledge, the heterogeneity of vocabulary and structure between KBs
makes sharing data between KBs difficult. 
What is needed is a KB equivalent of schema mapping and data exchange~\cite{fagin2005data} in which data structured under 
a source KB can be faithfully translated to a target KB.

\begin{figure*}
\centering
\includegraphics[width=1\textwidth]{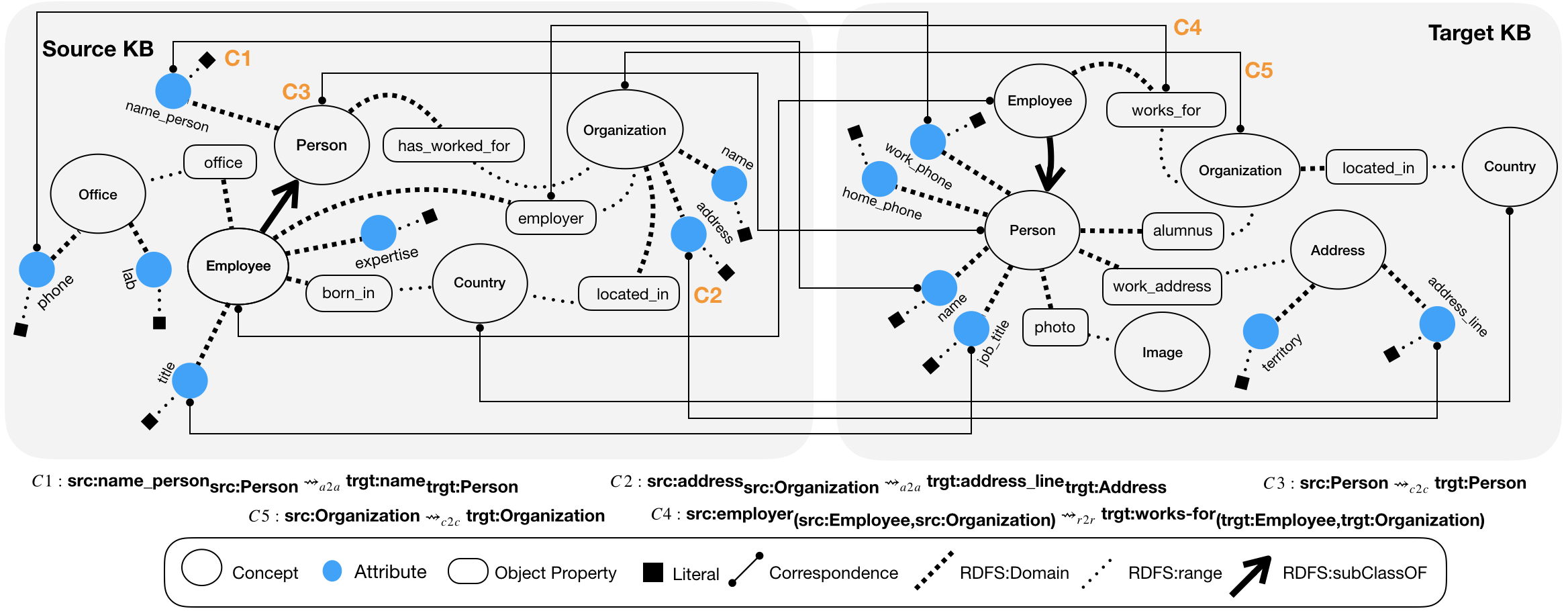}
\caption{\label{mainex}RDFS layer of two KBs and correspondences between them.}
\end{figure*}

Data exchange requires the existence of a set of rules (called mapping rules) that specify the relationship between the source and target.
It is important to note that even in the  relational model, heterogeneity cannot be reconciled with simple rules, called correspondences (or sometimes matches).  Correspondences  are created in ontology alignment (a.k.a. schema matching)~\cite{Euzenat:2013:OM:2560129} and ontology merging~\cite{stumme2001fca} techniques.  These  rules
only represent simple relationships (such as equivalence or containment) between small sets of resources.  This is all the more true in KBs,  as correspondences (even N:M correspondences) cannot express the complex relationships among many 
resources that need to be exchanged as a whole in order to preserve their relationships. 
Mapping in general requires complex logic or a full query language.   Duo et al.~\cite{dou2005ontology} argue that one of the \emph{biggest obstacles} to performing data exchange in KBs is the difficulty in manually creating mapping rules.
In traditional (non-KB) data exchange, there is a large body of literature on systems that reduce the burden of creating these rules manually, by making the rule creation process as automatic as possible.  See Bonifati et al.~\cite{bonifati2011discovery} for a survey on  mapping generation tools (or MGTs).  In comparison, there has been much less work on automatic mapping rule generation \emph{when the exchange is between two KBs}.  We refer to these tools as KMGT (knowledge-base mapping generation tools).
\paragraph*{Challenges and Contributions}
Several languages~\cite{alkhateeb2009extending,Bizer:2010:RFP:2878947.2878956,dou2005ontology,polleres2007sparql++} and frameworks~\cite{dou2005ontology,schultz2011ldif} have been proposed to help data engineers write KB mapping rules.   
In addition, there are a few pioneering KMGTs, including Mosto~\cite{rivero2011generating,rivero2013exchanging} and a system by Qin et al.~\cite{qin2007discovering} that automatically create KB mappings.  Here we describe 
some of the limitations of the current solutions (both KMGTs and MGTs) and describe our 
contributions. \\
\underline{\bf Associations:}
The main difference between \methodName and all other MGTs and KMGTs is in the way that it defines associations.
Data sharing tasks generally involve two steps: (1) correspondence creation (a.k.a alignment or matching) and (2) correspondence interpretation (a.k.a mapping creation). 
Central to step (2) is how to create associations -- the main semantic unit for associating resources so that they and their relationships are correctly mapped.  
In MGTs, 
tables
need to be combined to create associations.  Following the first relational MGT,  Clio~\cite{miller2000schema,Popa:2002:TWD:1287369.1287421}, most MGTs use declared (or discovered) constraints (like foreign keys or inclusion dependencies) to create mappings.
Using this approach, and excluding cycles, 
there are typically only a small number of ways  of joining any two tables. 
KBs are much more general graphs and contain large numbers of property paths.  In general, these property paths are not indicated as being full (functional) inclusions. 
This is an important property of KBs, their flexibility in representing information. 
\begin{example}\label{ex:associations}
Consider the source KB in Figure~\ref{mainex}.  In a relational world where \texttt{Organization} and \texttt{Country} are tables, if we know that if an \texttt{Organization} has value for the attribute \texttt{country} then it is  a valid key for \texttt{Country}  (i.e., there is a  FK from \texttt{Organization.country} to \texttt{Country}), then when mapping \texttt{Organization} and \texttt{Country} data to the target, MGTs will be sure to map an \texttt{Organization} with its own \texttt{country} (and not the country of a different organization).  However, if a \texttt{Person} has a \texttt{has-worked-for} attribute, but this attribute is not declared as 
having to have an organization (e.g., 
a FK constraint), then MGTs will not generate an association between Person and Organization.  In contrast, KBs often contain properties that hold only for portions of the data. The issue here is that KBs are  
open-world models and include  
many relationships
that exist only for a portion of the data.  Consequently, generating associations for all paths has been considered \emph{infeasible} by existing KMGTs~\cite{qin2007discovering,rivero2011generating,rivero2013exchanging}. \qed
\end{example}
To solve this complexity problem, Mosto~\cite{rivero2011generating,rivero2013exchanging} assumes that two concepts are associated only if they are connected via an aligned object property or if one is the ancestor of the other. 
Because of this assumption, Mosto \emph{relies heavily} on the existence of aligned
 properties, and if there are no correspondences between  
 properties, it cannot interpret correspondences collectively among concepts which are not subclass/superclass of each other. To mitigate  this, Mosto allows users to \emph {manually} add to the KB new constructs \texttt{mosto:strongRange} and \texttt{mosto:strongDomain} that {\em mimic} the role of a relational 
FKs. 
This permits Mosto to follow a MGT-like approach that  
also forms associations over paths that include these new, user-provided, annotations.  The newer MostoDex~\cite{rivero2015mostodex,rivero2016mapping}  can generate mapping rules without relying on 
any KB constructs, including user-provided annotations,
and instead  
uses user-provided examples. Both Mosto and MostoDex require a user to fully understand the nuances of KB translation in order to provide correct axioms or sufficiently informative examples.  In contrast, \methodName does not require user-provided KB constructs or user-provided examples.   To the best of our knowledge, \methodName is the first KMGT to consider associations that cover all property paths in both the source and target. It is worth mentioning that the KMGT proposed by Qin et al.~\cite{qin2007discovering} uses a similar approach to Mosto but does not allow user intervention.

\noindent \underline{\bf Structurally Valid Mappings:}
Considering all associations, however, can lead to an overwhelming number of mappings.  
To conquer this complexity (without eliminating desired mappings),  
we  define {\bf c2a valid} mappings, that eliminate many mapping alternatives by
requiring  mappings to respect the internal structure and interconnection of individuals and their attributes.
Importantly, mappings that are c2a valid ensure structural consistency of mapped resources even when value invention is required.  
In addition, \methodName uses correspondences between property paths when available, called r2r mappings, which may be provided by a 
matcher or correspondence generator,  to narrow down a set of good mappings
(called {\bf r2r valid} mappings).  
Note that previous approaches do not accept correspondences among object property \emph{paths} of length greater than one.  

\noindent \underline{\bf Knowledge Translation:}
The theory of data exchange (i.e. semantics and query answering) between KBs has already been studied~\cite{arenas2011knowledge,arenas2016knowledge,arenas2012exchanging,arenas2013data}.  However, the practicalities of generating queries that create valid solutions over real KBs has not received as much attention. \methodName's mapping rules are expressed using SPARQL queries which means they can be used directly to exchange knowledge from source to target.  Because we create associations using all paths in a KB, 
we have systematically considered how to create correct queries over associations that include property paths, without cardinality or functionality restrictions.  Our solutions 
perform the value invention needed (for example, when a target concept does not exist in the source, but its attributes do).  \methodName correctly associates data and translates all desired data (using the SPARQL optional command as required).
MGTs use sophisticated methods for value invention (using labeled nulls or skolem functions) such that the structure of the target and source are preserved~\cite{fagin2009clio,marnette2011++}. When dealing with KBs, \emph{blank nodes} fulfill this role. To the best of our knowledge, \methodName is the only KMGT that 
creates these blank nodes while preserving the grouping and relationships from the source KB. 

\begin{figure}
\centering
\centerline{
\includegraphics[width=0.45\textwidth]{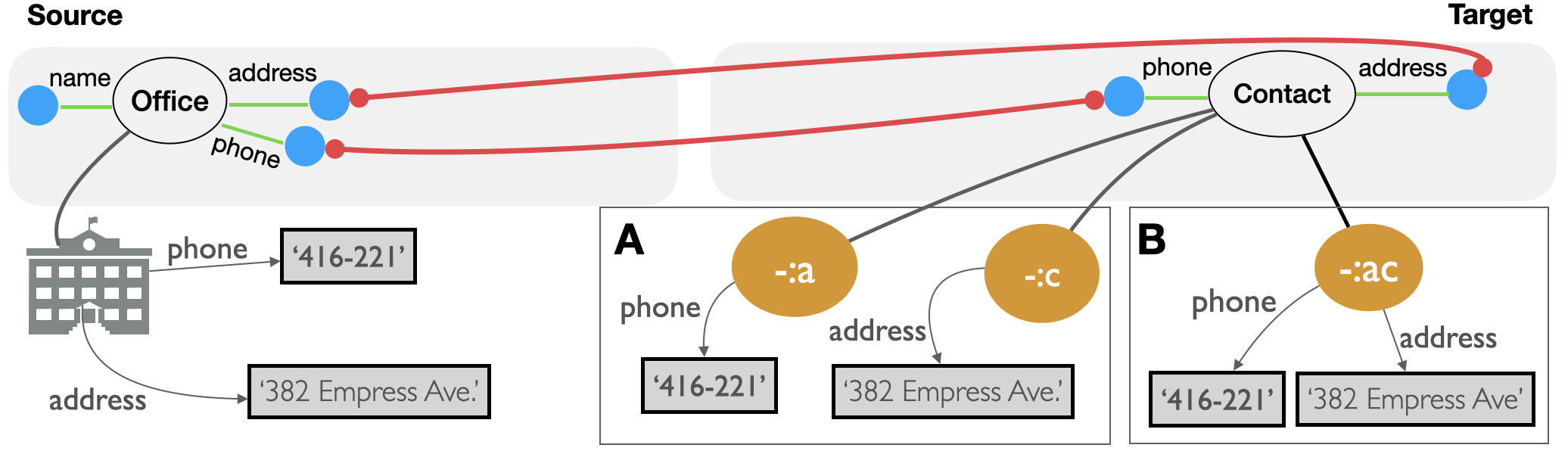}}
\caption{\label{ex2} Red lines represent a correspondence.}
\end{figure}

\begin{example}
Consider the scenario presented in Figure~\ref{ex2}. 
Current KMGTs create mapping rules which exchange data as shown in Box A.  They associate the phone number and address of a single source \texttt{Office} with two (possibly different) \texttt{Contact} resources in the target.  
This does not capture the semantic information encapsulated in the structure of the source KB, namely, that this phone and address are associated with one another. 
\methodName on the other hand, creates mapping rules that exchange data as shown in Box B, capturing the original grouping of data in the source KB.  Kensho can do this even if the source is incomplete (some offices do not have addresses, some do not have phones, or some have neither). 
\qed
\end{example}

\noindent\underline{\bf Ranking Heuristics:}
\methodName is the only tool that takes on the task of generating associations for all paths in the KB. While our structural validity requirements reduce the number of possible mappings, we may still generate many mappings.
To reduce the burden of selecting the best set of mapping rules for the exchange task at hand, we have proposed three new ranking heuristics to rank the final set of \emph{valid mappings}.
Buhmann et al.~\cite{buhmann2012universal} observe that, in practice, there is a lack of KBs that contain high quality schema axioms. 
Keeping this in mind, we have designed these heuristics such that they do not require KBs to both be populated with instances or to be annotated with reach axioms such as cardinality, functionality, or 
disjointness.

\noindent \underline{\bf Evaluation:}
We evaluated the performance of \methodName on several real-world scenarios designed to highlight the role of knowledge translation in different tasks including KB population, versioning, and migration. 
To show the performance of mapping generation in complex scenarios, 
we have also evaluated \methodName on a large number of synthetic scenarios. Our results show that \methodName  scales very well even for KBs that are three times larger than DBpedia in terms of the the number of concepts, with the largest bottleneck being the number of possible interpretations of a correspondence.  
We have also compared \methodName with existing KMGTs on three existing benchmarks.  
Finally, we have included a small case study to further investigate the effectiveness of \methodName.

\section{Methodology}

\methodName generates  executable mapping rules in two steps: \stepTwo and correspondence interpretation. When interpreted separately, correspondences cannot describe how to translate the resources of KBs  in conjunction with each other. To determine a set of executable mapping rules which weave these correspondences together, we must understand what relationships exist between aligned resources {\em within} each of the two KBs.  We call these relationships semantic associations. The goal of the first step 
is to discover these associations. The goal of the {\em correspondence interpretation} step is to interpret the set of given correspondences collectively in a way which respects the discovered semantic associations among elements of each KB. 
We begin by defining correspondences.

\subsection{Correspondences}\label{sec:prelim}

We distinguish between two important types of properties in a KB: 1)  a \emph{datatype property} or \emph{attribute} which represents a relationship between an IRI and a Literal 
and 2) an \emph{object property} 
which expresses a relationship between two IRIs.
A property path is a 
list of properties in
an RDF graph between resources~\cite{sparql2013}.
An 
attribute or an object property is a property path of length one.
Property paths are either data property paths (a path between an IRI and a Literal) or object property paths (a path between two IRIs).
Property paths can be expressed using a regular expression grammar.  
We use the 
grammar from W3C~\cite{sparql2013} to represent them. 

In this work, we  
define three  types of correspondences. First, a {\bf Concept2Concept} correspondence 
associates
a concept $\texttt{s}$ in the source to a concept $\texttt{t}$ in the target, represented as $\texttt{s} \leadsto_{c2c} \texttt{t}$. 
In Figure~\ref{mainex}, $C3$ is a Concept2Concept correspondence between the source concept \texttt{Person} and the target concept \texttt{Person}.
Second, a {\bf Rel2Rel} correspondence  associates an object property path \texttt{P} in the source to an object property path \texttt{R} in the target, represented as $\texttt{P}_{(\texttt{s}_1, \texttt{s}_2)} \leadsto_{r2r} \texttt{R}_{(\texttt{t}_1,\texttt{t}_2)}$ where \texttt{$\texttt{s}_1$} and \texttt{$\texttt{s}_2$} are connected by \texttt{P}  and  \texttt{$\texttt{t}_1$} and \texttt{$\texttt{t}_2$} are connected by \texttt{R}. 
In Figure~\ref{mainex}, $C4$ is a Rel2Rel correspondence that associates the source object property \texttt{employer} with the target object property \texttt{works\_for}.
An {\bf Attr2Attr} correspondence associates a data property path 
in the source with a data property path 
in the target, represented as $\texttt{a}_{{\texttt{s}}} \leadsto_{a2a} \texttt{b}_{{\texttt{t}}}$, where \texttt{$\texttt{s}$} and \texttt{$\texttt{t}$} are concepts 
that are connected by a data path to attribute values of $\texttt{a}$ and $\texttt{b}$, respectively. 
In Figure~\ref{mainex}, $C2$ is an Attr2Attr correspondence that associates the source attribute \texttt{address} with the target attribute \texttt{address\_line}.
Correspondences can represent various relationships. 
In this work, we use correspondences that express subset-or-equal relations since correspondences produced by automated tools are often of this type ~\cite{Euzenat:2013:OM:2560129}. Following the terminology of Euzenat and Shvaiko~\cite{Euzenat:2013:OM:2560129}, we call a \textbf{set} of correspondences an \emph{alignment}. 

We say a concept is an \emph{aligned concept} if it either directly participates in a Concept2Concept correspondence or if it participates in  an Attr2Attr correspondence (meaning for $\texttt{a}_{{\texttt{s}}} \leadsto_{a2a} \texttt{b}_{{\texttt{t}}}$, it is the concept \texttt{s} or some concept along the path from \texttt{s} to the attribute \texttt{a}).   Note that we treat Rel2Rel correspondences differently, and use them to refine how concepts are mapped collectively using a notion we call {\bf r2r validity} defined in the next section.  
Finally, in general we call any KB resource that participates in a correspondence an \emph{aligned element}. It is worth mentioning that Mosto also uses another type of correspondence that matches an attribute value to an instance of a concept. In these situations, we create an Atr2Atr that matches the attribute in the source to the \texttt{rdfs:label} attribute of the concept in the target. 
Note that this type of correspondence is different from metadata-data correspondences~\cite{hernandez2008data,Mil98} which \methodName does not support.

\subsection{Semantic Association Discovery}\label{SAD}\label{sec:semanticassociation}
In a KB, semantic associations between 
aligned elements
are represented as property paths between them. 
First, a concept and its 
attributes indicate a semantic association (sometimes called the internal structure of the KB~\cite{Euzenat:2013:OM:2560129}). 
In Figure~\ref{mainex},
the fact that \texttt{trgt:name} and \texttt{trgt:work\_phone} are 
attributes
of the concept \texttt{trgt:Person} indicates that for each individual of type \texttt{trgt:Person}, the value of these 
attributes
are semantically related. 
Second, two concepts (and their attributes) are associated when there is a property path between them. 
For instance, information about an organization and its employees may be modeled as a 
path containing a number of concepts.
Of course, in the presence of cycles, the number of associations is infinite so \methodName will only enumerate a finite set of these.  

\begin{definition}{(Basic Association)}\label{def:ba}
Each aligned concept defines a Basic Association that includes the concept along with all its aligned attributes.
We call the aligned concept
the \emph{root} of the basic association.
\qed  
\end{definition}

In addition to basic associations, aligned concepts can be associated by the
relational~\cite{Euzenat:2013:OM:2560129} structure of a KB.
To 
define
these associations, we 
use
property paths between the roots of basic associations.

\begin{example}\label{ex:externalStruct}
In Figure~\ref{mainex}, the source concepts \texttt{Organization} and \texttt{Country} are associated through the path \texttt{located\_in}. The instances of this \emph{association} can be retrieved using the following two queries.

\noindent\begin{minipage}[t]{.5\linewidth}
\vspace{-2mm}
{\obeylines\obeyspaces\scriptsize
\texttt{
\begin{linenumbers}SELECT * WHERE \{
    ?o a src:Organization.
    OPTIONAL 
    ~~~\{?o  src:located\_in  ?ozCountry.
    ~~~~?ozCountry  a src:Country.\}\}
    \end{linenumbers}
}} 

\end{minipage}%
\begin{minipage}{.1\linewidth}
\vspace{-2mm}
\end{minipage}%
\begin{minipage}[t]{.5\linewidth}
\vspace{-2mm}
{\obeylines\obeyspaces\scriptsize
\texttt{
\begin{linenumbers}SELECT * WHERE \{
    ?c a src:Country.
    OPTIONAL 
    ~\{?c  $\texttt{src:located\_in}^{\wedge}$  ?czOrganization.
    ~~?czOrganization  a src:Organization.\}\}
    \end{linenumbers}
}}
\medskip \resetlinenumber
\end{minipage}

\noindent The property $\texttt{src:located}\_\texttt{in}$  models the relationship between the
two 
aligned concepts
\texttt{Organization} and \texttt{Country} which are the roots of basic associations. 

Note that in KBs,
the source might be incomplete,  
hence, the query needs to contain the \texttt{OPTIONAL} keyword.
\qed 
\end{example}

\noindent To directly associate two aligned concepts, we use paths, called {\em association paths} that do not go through other aligned concepts. We limit the length of these paths to $\cal D$.
To simplify notation,  we assume that each instance in our KBs has exactly one most-specific type. 
This is not an essential assumption. In \TRep{Section~\ref{multiInherit}}, 
we briefly discuss how this assumption may be relaxed by
generalizing the definition of Association Path.

\theoremstyle{definition}
\begin{definition}{(Association Path)}\label{def: associationPath}
An association path $p$ 
between aligned concepts $u_0$ and $u_1$ is an ordered 
list of resources
on object property path $\Pi$ which matches:
($\texttt{rdfs:domain}^{\wedge}|\texttt{rdfs:subClassOf}$)/ $(\texttt{rdfs:domain}^{\wedge}|$\\
\rightline{$\texttt{rdfs:domain}|\texttt{rdfs:range}|\texttt{rdfs:range}^{\wedge}|\texttt{rdfs:subClassOf})^*$}\\
such that $(u_0,u_1)$ is in the evaluation of $\Pi$, and there is no aligned concept along $\Pi$.
The concept $u_1$ (also called \emph{tail} of the association path) is part of the path, but $u_0$ (also called the \emph{root} of the association path) is not part of the path.
\qed
\end{definition}

\noindent 
\TRep{Intuitively, Basic associations can be related using {\em association paths} which are paths in the KB from an aligned concept $u$ to another aligned concept $v$ where the path does not contain any other aligned concept.  
each association path $\cal P$ 
between $u$ and $v$ is an ordered 
list of resources on a property path that matches the expression $R/Q$. 
In $R/Q$, $R$ is a property path that is either an inverse domain (so $u$ is the domain) or $u$ is a subclass.  
{That is, $R$ matches $\texttt{rdfs:domain}^{\wedge}|\texttt{rdfs:subClassOf}$}.\\
To $R$, we \emph{may} concatenate a path $Q$ that contains a range (leading to another aligned concept $v$) 
or a longer path that goes through other, non-aligned concepts and eventually reaches $v$. Specifically, $Q$ matches:\\
\centerline{$(\texttt{rdfs:domain}^{\wedge}|\texttt{rdfs:domain}|\texttt{rdfs:range}|\texttt{rdfs:range}^{\wedge}|\texttt{rdfs:subClassOf})^*$}
}
\begin{example}\label{ex:associationPath} 
In Figure~\ref{mainex}, there is only one association path that connects the target concept \texttt{Organization} to \texttt{Country}, namely \texttt{[located\_in, Country]}.  Note there are no association paths from \texttt{Country} to \texttt{Organization} (because $R$ does not include $\texttt{rdfs:range}^{\wedge}$).
This is simply to prohibit enumerating redundant paths. 
If we consider \texttt{Employee} to \texttt{Person} (again in the target), then 
\texttt{[Person]} is an association path which is created by going through the \texttt{rdfs:subClassOf} property path from \texttt{Employee}.  If \texttt{Organization} were {\em not} aligned then 
$\texttt{[works\_for, Organization, alumnus}^{\wedge} \texttt{, Person]}$
would also be an association path. \qed
\end{example}

\TRep{In creating association paths, we will not allow backtracking. That is, if node $B$ can be reached using property $p$ from node $A$, in the next step, $p^{\wedge}$ will not be traversed \emph{unless} the object of $p^{\wedge}$ is something other than $A$. For instance, in Figure~\ref{mainex}, $\texttt{[photo, Person]}$ is not a valid association path from \texttt{trgt:Person}, since in order to create it we need to follow $\texttt{rdfs:domain}^{\wedge}\texttt{/rdfs:domain}$. However, for instance in Figure~\ref{ex1}, in the source KB, $\texttt{[hasSupervisor, Person]}$ is a valid association path from \texttt{Person}, since it is created by following $\texttt{rdfs:domain}^{\wedge}\texttt{/rdfs:range}$. One consequence of this strategy is that an attribute will never be added to an association path, and thus elements of an association path can only be concepts and object properties.}

We use association paths to connect basic associations.
Semantic associations start with a basic association for an aligned concept $u_0$.  We add to $u_0$ all association paths to other aligned concepts $u_i$ together with the basic association for $u_i$.  We repeat this process recursively adding to the semantic association paths from $u_i$ to
other aligned concepts $u_j$ together with the basic association of $u_j$.  We limit the length of the longest path in the semantic association to \length.

\theoremstyle{definition}\label{sa} 
\begin{definition}{(Semantic Association)}
The semantic association for aligned concept $u_0$, denoted $\sa{u_0}$, contains the basic association for $u_0$.  In addition, if $u_i$ is a concept in 
$\sa{u_0}$ and there is an association path $p$ from $u_i$ to aligned concept $u_j$ and if adding $p$ to $\sa{u_0}$ does not create a path longer than length \length\ in $\sa{u_0}$, then the association path $p$ is in $\sa{u_0}$ and the basic association of $u_j$ is also in $\sa{u_0}$.
For each concept in each $p$ added to $\sa{u_0}$, a new node will be created (even if a node representing that concept is already in $\sa{u_0}$). 
A semantic association
  $\sa{u_0}$ can be 
thought
of as a  tree, 
where the concepts in $\sa{u_0}$ are nodes and the root is a node representing concept $u_0$. We define this tree such that if $u_i$ and $u_j$ are associated via a \texttt{rdfs:subclassOf} path, then there is an edge between them labeled with \texttt{a}. Otherwise, the edge is labeled with the corresponding property.
Each node $n \in \sa{u_0}$ is assigned a variable (denoted $var(n)$) and we use the notation $concept(n)$ to denote the concept that $n$ represents. Also, if the basic association of $concept(n)$ is in $\sa{u_0}$, each attribute $a$ in this basic association is assigned a  variable (denoted $var(n,a)$). For a semantic association $\cal A$, we call the set of all variables assigned to its nodes and attributes, denote ${\cal A}_{vars}$, the \emph{variables} of the semantic association.
\qed
\end{definition}

\begin{figure*}
\centering
\includegraphics[width=0.8\textwidth]{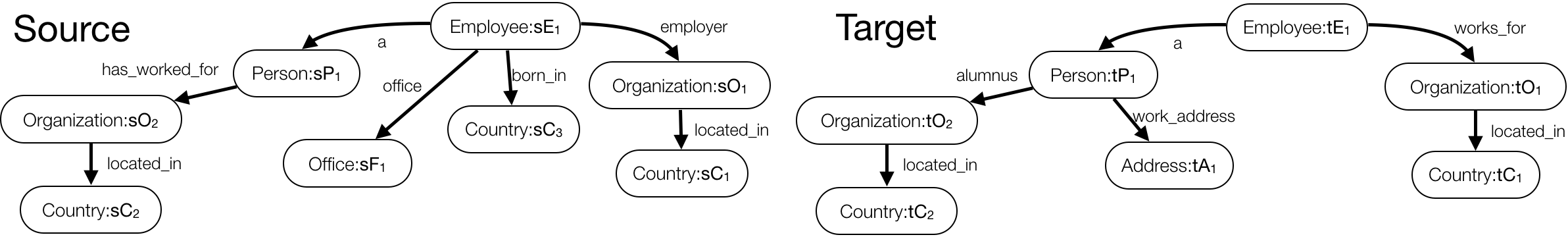}
\caption{\label{fig:saEmp}{
Semantic associations \sa{\texttt{src:Employee}} and \sa{\texttt{trgt:Employee}} with attribute variables omitted.
}}
\end{figure*}

For the KBs of Figure~\ref{mainex}, the semantic associations of the source and target \texttt{Employee} concepts are depicted in Figure~\ref{fig:saEmp}.
Note that cycles and KBs with multiple property paths between the same concepts can lead to semantic associations containing multiple nodes for the same concept.  

\subsection{From Correspondence to Mapping}
\label{sec:interp}
Semantic associations (within the source and target) can be used to understand correspondences collectively.
\begin{example}\label{constructQ}

The correspondences $C1$ and $C2$ of Figure~\ref{mainex} can be
interpreted independently. If we do this, we would get mappings that can be represented by the following 
queries.

\noindent\begin{minipage}[t]{.46\linewidth}
{\obeylines\obeyspaces\scriptsize
\vspace{-2mm}
\texttt{
\begin{linenumbers}
\underline{construct}\{
~-:trgtPerson a trgt:Person.
~-:trgtPerson trgt:name \textbf{?trgtName.}\}
\underline{where}\{
~?srcPerson a src:Person.
~?srcPerson src:name\_person \textbf{?srcName}.
~bind(\textbf{?srcName} as \textbf{?trgtName})\}
    \end{linenumbers}\resetlinenumber
}} 
\medskip
\end{minipage}%
\begin{minipage}{.1\linewidth}

\end{minipage}%
\begin{minipage}[t]{.6\linewidth}
{\obeylines\obeyspaces\scriptsize
\vspace{-1.5mm}
\texttt{
\begin{linenumbers}
\underline{construct} \{   
~-:tAddress a trgt:Address.
~-:tAddress trgt:address\_line \textbf{?trgtAddress}.\}
\underline{where} \{?srcPerson a src:Person. 
~?srcPerson src:has\_worked\_for ?srcOrg.
~?srcOrg src:address \textbf{?srcAddress}.
~bind(\textbf{?srcAddress} as \textbf{?trgtAddress})\}
  \end{linenumbers}\resetlinenumber
}}
\medskip
\end{minipage}

\noindent These mappings create target \texttt{addresses} from source \texttt{addresses} (and target \texttt{names} from source \texttt{names}), but do not associate \texttt{names} and
\texttt{addresses} in the target. 
To preserve source information, specifically the association between 
resources, we need to understand when a set of correspondences can be interpreted collectively.
For instance, 
if $C1,C2,$ and $C3$ 
were the only correspondences in Figure~\ref{mainex}, then the following mapping
is a better interpretation of the correspondences.

{\obeylines\obeyspaces\scriptsize
\texttt{
\begin{linenumbers}
~~\underline{construct} \{\textbf{?trgtPerson} a trgt:Person.
~~~~~ ?trgtPerson trgt:name \textbf{?trgtName}.
~~~~~ ?trgtPerson trgt:work\_address -:trgtAddress.
~~~~~ -:tAddress a trgt:Address.
~~~~~ -:tAddress trgt:address\_line \textbf{?trgtAddress}.
~~~~~\}
~~\underline{where} \{\textbf{?srcPerson} a src:Person, 
~~~~~~~~OPTIONAL\{?srcPerson src:name\_person \textbf{?srcName}\}
~~~~~~~~OPTIONAL\{?srcPerson src:has\_worked\_for ?srcOrg
~~~~~~~~~~~~OPTIONAL\{?srcOrg src:address \textbf{?srcAddress}\}\}
~~~~~~~~bind(\textbf{?srcName} as \textbf{?trgtName})
~~~~~~~~bind(\textbf{?srcAddress} as \textbf{?trgtAddress})
~~~~~~~~bind(\textbf{?srcPerson} as \textbf{?trgtPerson})\}
  \end{linenumbers}\resetlinenumber
}}
\medskip

Similar to the mappings that interpret each correspondence independently, the above mapping dictates how to create target \texttt{addresses}, \texttt{names}, and \texttt{persons} from the resources of the source. However, 
this mapping \emph{also} preserves the relationships among the translated
elements, and thus is usually more desirable. 
\qed
\end{example}

The first step of query generation is to find correspondences that can be interpreted together.
To do this, we consider pairs containing one source semantic association and one target semantic association and define the set of correspondences that are {\em covered} by this pair (e.g., the pair $\langle$\sa{\texttt{src:Person}}, \sa{\texttt{trgt:Person}}$\rangle$ covers the three correspondences in the above example).  
Following the terminology used in Clio~\cite{Popa:2002:TWD:1287369.1287421}, a 
\emph{Skeleton} is a pair $\langle S,T \rangle$ where $S$ is a source
{\finalAssociation}  and $T$ is a target {\finalAssociation}.
To identify how a correspondence $C$ can be interpreted using a skeleton, or in other words to identify the coverage of correspondence $C$ by a skeleton,  it is not enough to check whether the {\finalAssociation}s include the 
aligned elements which are participating in the correspondence $C$. One reason is that the same concept of the KB might be included more than once in a {\finalAssociation}.
Hence, we define a \emph{renaming function} that associates target variables of the target association to  variables of the source association.  We do this in a way that respects the correspondences. 
In general, there may be multiple ways to cover a correspondence with respect to a pair of source and target {\finalAssociation}s 
and each represents a specific (different) interpretation of the correspondences.
We give an example, then formally define coverage.

\begin{example}\label{ex:multi-cover}
Figure~\ref{fig:saEmp} depicts a pair of source and target semantic associations.  
All nine correspondences shown in Figure~\ref{mainex} are covered by this pair of semantic associations.  
Some of the correspondences such as $\texttt{src:Country} \leadsto_{c2c} \texttt{trgt:Country}$ can be covered in multiple ways.
Intuitively, there are six possible interpretations of the country
correspondence.  
The country of the organization where an employee works  (target variable $tC_1$) can be populated with the country of an organization which is her employer ($sC_1$), or the country of an organization for which she has worked ($sC_2$), or with the country where she was born ($sC_3$).
Similarly, the country of the organization of which a target employee is an alumni (variable $tC_2$) can be populated with any of these three source options.  
Note that there are other options, for example, one might
decide to map both target countries to source data, or
to leave
one of these target countries unmapped.
Also note that we are mapping each employee with a country (maintaining this association), so some of these options may require value invention for concepts along a target path that do not exist in the source.  
\qed
\end{example}

We now define this formally by defining renamings
from the variables of $T$ (denoted by $T_{vars}$), to the variables of $S$ (denoted by $S_{vars}$).  
A renaming is a \emph{total function} that maps each variable of $T$ either to a variable of the source or to $\epsilon$ which will be an indication that the mapping query \emph{might} need to perform value-invention for this variable (something we discuss in Section~\ref{sec:mapgen}).  

\begin{definition}{(Correspondence Coverage)}
A correspondence $C$ is {\em covered} by a skeleton $\langle S,T \rangle$ if there is a \emph{renaming} 
$\Re: T_{vars} \rightarrow S_{vars} \cup \{\epsilon\}$, where:

\squishlist
\item 
(Concept2Concept) If $C: s \leadsto_{c2c} t$, then $\exists n \in S, m \in T$,  $\Re(var(m)) = var(n)$,  $concept(n) = s$, $concept(m) = t$.  
\item
(Rel2Rel) If $C: P_{(s_1, s_2)} \leadsto_{r2r} R_{(t_1,t_2)}$,
then $\exists n_1, n_2 \in S$ 
and $\exists m_1, m_2 \in T$, 
$\Re(var(m_1)) = var(n_1)$,
$\Re(var(m_2)) = var(n_2)$, 
 $concept(n_1) = s_1$, 
 $concept(n_2) = s_2$, 
$concept(m_1) = t_1$, and 
$concept(m_2) = t_2$. 
\item 
(Attr2Attr) If $C: a_s \leadsto_{a2a} b_t$, 
then $\exists n \in S, m \in T$,  
 $concept(n) = s$, $concept(m) = t$,
$\Re(var(m,b)) = var(n,a)$.

\squishend
If a correspondence $C$ is covered by $\langle S,T \rangle$ using a renaming $\Re$, then we call $\Re$ an {\em interpretation} of $C$. 
\qed
\end{definition}

\noindent Notice that a single renaming may be an interpretation for many correspondences.

\begin{definition}{(Skeleton Renaming)}
\label{def:skeletonrenaming}
Given a skeleton $\langle S,T \rangle$ and a set of correspondences $\cal C$, then $\Re$ is a renaming for $\cal C$ if it includes at least one 
interpretation
of each correspondence in $\cal C$. 
The set  $\Bar{\Re}_{S,T}$ denotes all possible $\Re$s over
a skeleton
$\langle S,T \rangle$. 
\qed
\end{definition}

\begin{figure}
\centering
\includegraphics[width=0.45\textwidth]{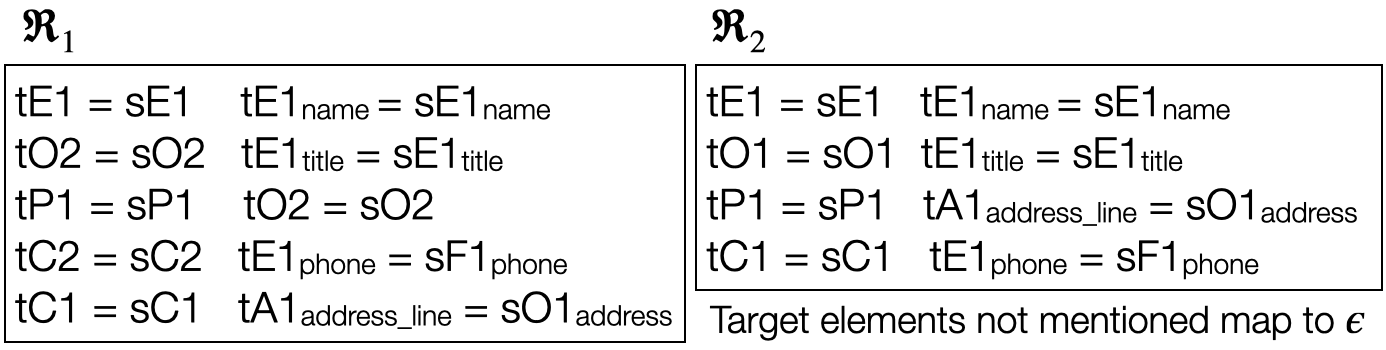}
\caption{\label{Re} Two of the possible renamings of $\langle \sa{\texttt{src:Employee}},\sa{\texttt{trgt:Employee}} \rangle$.}
\end{figure}

\begin{example}\label{coverage}
In Figure~\ref{mainex}, 
let $\cal C$ contain all correspondences except $C4$.
Figure~\ref{fig:saEmp} depicts source and target associations
for the skeleton $\langle \sa{\texttt{src:Employee}},\sa{\texttt{trgt:Employee}} \rangle$. Given $\cal C$,
Figure~\ref{Re} shows two possible renamings ($\Re_1$ and $\Re_2$) for this skeleton.
In comparison with $\Re_1$, 
$\Re_2$ does not use the source variables 
$\texttt{sO2}$ and  
$\texttt{sC2}$, and the target variables $\texttt{tO2}$ and $\texttt{tC2}$ now map to $\epsilon$.
The renaming $\Re_2$ might be desirable if a data engineer decides that the relationship \texttt{trgt:alumnus} between \texttt{trgt:Person} and \texttt{trgt:Organization} in the target is not expressed by any property path between \texttt{src:Person} and \texttt{src:Organization} in the source, and thus should not be part of the translation.\qed
\end{example}

\noindent To accommodate the modeling flexibility of KBs, we have defined
  semantic associations to include any paths up to a given length (not just functional, total, or
  user indicated paths).  Thus, there can be a large number of possible
  renamings 
  for a pair of semantic associations.  
  We define three notions of validity, each of which reduces the
  number of possible 
  renamings without excluding desirable renamings.
  
The first is similar to what is used now by existing KMGTs~\cite{qin2007discovering,rivero2011generating,rivero2013exchanging}
  and requires that the renaming only map concepts and
 attributes for which there is a correspondence (and map them as
 indicated by the correspondence).  The second requires the 
  renaming to respect all correspondences between relationships
  (property paths).  Note that the set of $Rel2Rel$ correspondences may be
  incomplete (due to limitations of current alignment tools), but if present, the renaming must respect them. 
 The third requires the renaming
    to respect the internal structure of the KB and interconnection
    between concepts and attributes, and map the attribute values to
    individuals correctly.

\begin{definition}{(Baseline Validity)}\label{basicVal}
Given a set of correspondences $\cal C$ and skeleton $\langle S,T \rangle$, renaming $\Re$ is {\basic} valid if:
\squishlist
\item 
$\forall n \in S, m \in T$,  if $\Re(var(m)) = var(n)$,  $concept(n) =
s$, and $concept(m) = t$, then there is a correspondence  
$s \leadsto_{c2c} t  \in \cal C$   
\item
$\forall n \in S, m \in T$,  if  
$\Re(var(m,b)) = var(n,a)$,
$concept(n) = s$, $concept(m) = t$, then $\exists$  
$a_s \leadsto_{a2a} b_t  \in \cal C$ \qed
\squishend
\end{definition}

\noindent  
The number of baseline valid renamings can still be large.
\methodName use path correspondences (Rel2Rel) to narrow down the set of possible valid renamings. 
\begin{figure}
\centering
\includegraphics[width=0.45\textwidth]{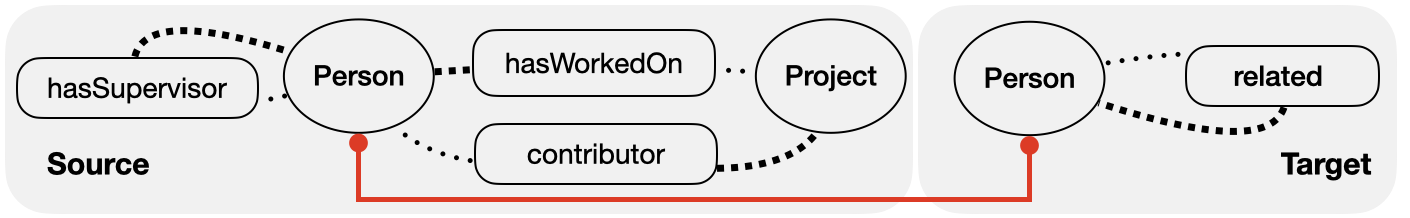}
\caption{\label{ex1}
Red line represents a correspondence.
}
\end{figure}

\begin{definition}{({\rel} Validity)}\label{relVal}
Given a Rel2Rel correspondence $C: P_{(s_1, s_2)}
\leadsto_{r2r} R_{(t_1,t_2)}$ and skeleton $\langle S,T \rangle$, a renaming $\Re$ is {\rel} valid for $C$ if: $\exists m_1, m_2 \in T$, and
$\exists n_1, n_2 \in S$,
where $concept(m_1) = t_1$ and $concept(m_2) = t_2$ 
and $concept(n_1) = s_1$ and $concept(n_2) = s_2$, and
$\Re(var(m_1)) = var(n_1)$ and 
$\Re(var(m_2)) = var(n_2)$, and $n_1$ and $n_2$ are connected through $P$ in $S$, and $m_1$ and $m_2$ are connected through $R$ in $T$.  
\qed
\end{definition}
\noindent Previous KMGTs do not take advantage of Rel2Rel correspondences in which the length of the corresponding property paths is greater than one.  
For instance in Figure~\ref{ex1}, none of the current KMGTs 
consider a correspondence that represents the fact that two persons who are working on a project in the source are related in the target since this requires mapping
the \texttt{hasWorkedOn/contributer} property path in the source to the \texttt{related} property in the target.
Note this example requires no value invention and yet is typically not considered in the literature.  In addition, \methodName considers path correspondences where not every resource on the target path exists in the source, and hence value invention is required. (See Section~\ref{sec:mapgen} for more detail.)

An important innovation in \methodName is to consider the different ways in which each correspondence can be covered in a mapping.
Our associations 
can capture multiple interpretations of the same correspondence even in a single renaming (see Example ~\ref{ex:multi-cover}). This functionality is important especially in cases where value invention is needed, since 
correct mapping requires that
all desired resources and their relationships must be in a single query so that the newly created blank node IRIs can group resources properly. However, by allowing multiple interpretations of a correspondence, a renaming might need to map multiple attributes or concepts of the same type. For instance, in Figure~\ref{mapping} our association needs to express a rule that describes an employee, her current employer address, and the address of places where she has worked in the past. In such cases, we need to make sure that the addresses are mapped to the correct organizations.
\methodName   uses {\atr} validity to ensure concepts
and attributes are collectively mapped properly.
The main goal of {\atr} validity
is to require the mappings to respect the
internal structure of the source KB and interconnections between concepts and
attributes, so that the attribute values of individuals are preserved. 

\begin{definition}{({\atr} Validity)}\label{c2aVal}
Given two correspondences
$C_1: s \leadsto_{c2c} t$ and
$C_2: a_s \leadsto_{a2a} b_t$, and 
a skeleton $\langle S,T \rangle$, a renaming $\Re$ is {\atr} valid  for $C_1$ and $C_2$ if: 
$\forall n \in S, m \in T$, such that $concept(n) = s$, $concept(m) = t$, if $\Re(var(m)) = var(n)$, then $\Re(var(m,b)) = var(n,a)$.
\qed
\end{definition}

\noindent 
Note that a data engineer might choose a renaming that does not transfer the attribute value(s) at all, but if the value 
is mapped, then that value should be associated with the same individual which is mapped from the source. Although current KMGTs do not consider this validity,  traditional mapping generation tools (MGTs) produce mappings which are {\atr} valid. 
\TRep{In addition to the validity constraints mentioned above, we have used other constraints to further refine our renamings. One of these constraints states that in case that the domain of the attribute in the target is of type which only its attributes participate in the correspondences, those attributes can only be mapped to attribute values of the individuals in the source which are transferred to the target (of course, if the individual is of type which has a corresponding element in the target.) Another constraints makes sure that the root variable is mapped properly. Note that the variable that represents the root of \sa{r} can only be mapped to few certain variables, otherwise we will produce a large number of redundant mappings. Our algorithm ensures that only variables of the source which have the same type as $concept(r)$ and are in the minimum distance from the source's semantic association's root are mapped to the root variable in the target.}

\TRep{It is worth mentioning that not all skeletons produced using our algorithm can be used to interpret correspondences. Skeletons that do not cover any correspondences, such as 
$\langle \sa{\texttt{src:Country}},\sa{\texttt{trgt:Address}} \rangle$
cannot be used to generate any interpretations. In addition, sometimes a set of skeletons produce the exact same interpretations, and thus we should use only one of these skeletons to reduce the redundancy.
\methodName identifies redundant skeletons and eliminates them (this phase is called \emph{skeleton pruning}). Skeleton $\langle S,T \rangle$ is redundant if there exist another skeleton $\langle S',T' \rangle$ where the coverage of $\langle S',T' \rangle$ is the same as $\langle S,T \rangle$, and $|T'_{vars}| \leq |T_{vars}|$, where $T_{vars}$ and $T'_{vars}$ represent the set of all variables of $T$ and $T'$ respectively, and $S'$ is the same semantic association as $S$. For instance, $\langle \sa{\texttt{src:Country}},\sa{\texttt{trgt:Person}} \rangle$ and  $\langle \sa{\texttt{src:Country}},\sa{\texttt{trgt:Country}} \rangle$  both only cover { $\texttt{src:Country} \leadsto_{c2c} \texttt{trgt:Country}$}, however in the first skeleton $\sa{\texttt{trgt:Person}}$ represents variables other than country which will not be used during the translation. One of these skeletons is redundant and needs to be removed. In this case, between these two skeletons, \methodName only considers the simpler skeleton, i.e $\langle \sa{\texttt{src:Country}},\sa{\texttt{trgt:Country}} \rangle$, which has fewer or equal number of variables in the target's semantic association and will eliminate $\langle \sa{\texttt{src:Country}},\sa{\texttt{trgt:Person}} \rangle$.}

\subsection{Mapping Generation}
\label{sec:mapgen}
\begin{figure*}
\centering
\includegraphics[width=1\textwidth]{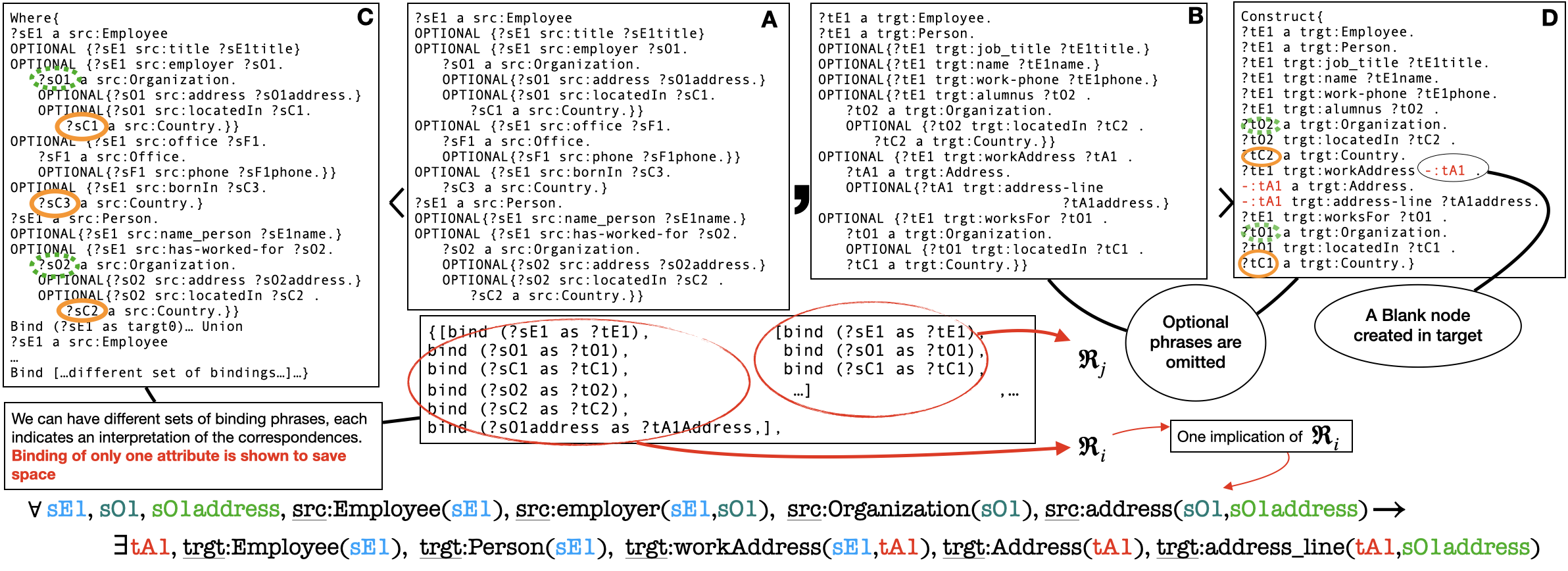}
\caption{\label{mapping}{ 
Query generation process. 
}}
\end{figure*}

For each skeleton renaming $(\langle S,T \rangle, \cal C, \Re)$, we can create a mapping.  We use the semantic associations $S$ and $T$ to create source and target 
query patterns.  Then we use $\Re$ to create what is effectively an inclusion dependency
from the source query pattern to the target query pattern.
We first 
describe how an 
{\bf association query pattern} is defined based on a semantic association. 
Given a semantic association $\sa{u_0}$, to create an association query pattern, we start from the root node $n_0$ 
(where $concept(n_0)$ is $u_0$). 
We use $var(n_0)$, $v_0$, to create a SPARQL pattern that expresses the type of instances of node $n_0$. That is, $\texttt{?v}_0 \texttt{ a u}_0$.  
For each of the root's children, $n_i$, we create a fact that represents $l(u_0,concept(n_i))$, where $l$ is the label of the edge which connects $n_0$ to $n_i$. That is, $\texttt{?v}_0 \texttt{ l ?v}_i$, where $v_i$ is $var(n_i)$ unless $l$ is label \texttt{a}, in that case, the variable of the parent, $v_0$, will be used instead for representing that node through out the whole process of query generation. We repeat the process recursively for each of $n_0$'s children. If $l$ is anything other than label \texttt{a}, we will nest what follows for that child in the \texttt{OPTIONAL} clause. Also, for each node, $n_j, j\geq 0$, we will express each of $concept(n_j)$'s attributes $q$, using statements like $\texttt{?v}_j \texttt{ q ?attr}_j$, where $\texttt{attr}_j$ is $var(n_j,q)$, and $v_j$ is $var(n_j)$. Each of these attribute statements will be nested within an \texttt{OPTIONAL} clause.

\begin{example}
Figure~\ref{mapping} Box A and B, are association query patterns created from the semantic association trees of $\sa{\texttt{src:Employee}}$  and $\sa{\texttt{trgt:Employee}}$ of Figure~\ref{fig:saEmp} respectively. 
Note that in both patterns, the \texttt{Employee} variable is used for representing the \texttt{Person} - and the clause that represents the fact that an employee is a person is not nested within an \texttt{OPTIONAL} keyword. The reason is that the semantics of inheritance implies that if \texttt{B} is a subclass of \texttt{A}, then every individual of type \texttt{B} is also of type \texttt{A}. \qed
\end{example}

Given source and target association query patterns generated from a skeleton $\langle S,T \rangle$, we now consider how skeleton renamings can be used over association query patterns to create mappings.
We create a SPARQL \emph{construct query} for each skeleton. 
Assuming that there are $N$ possible renamings for skeleton $\langle S,T \rangle$,
(that is, $|\Bar{\Re}_{S,T}| = N$), \textit{for each} renaming
$\Re \in \Bar{\Re}_{S,T}$, 
an association query pattern of
$S$ 
will be expanded with a set of binding clauses created from $\Re$ to create an $S_{\Re}$ graph pattern.  
More specifically, for each target variable $v_i$ which is mapped to a source variable $\Re(v_i)$, the binding clause ``\texttt{bind $\Re(v_i)$ as $v_i$}'' will be created and added to the triple pattern of $S_{\Re}$.
The final \underline{\texttt{Where}} clause of the construct query will be the \emph{union} of all 
graph patterns
(see Figure~\ref{mapping} Box C for an example.) 
Note that a data engineer can choose to only use a subset of 
the renamings in $\Bar{\Re}_{S,T}$.
In particular, in practice, we expect to only be working with valid renamings.

In order to create the \underline{\texttt{construct}} clause of our mapping query, it is important to note that 
some variables of the target's {\finalAssociation} may be mapped to $\epsilon$.
In order to materialize the target, sometimes we will have to fill in the values for the undetermined variables.
For instance, the renaming $\Re_2$ of Example~\ref{coverage}, shown in 
Figure~\ref{Re}, represents a translation which transfers the attribute values of  $sO1_{address}$ to  $tA1_{address\_line}$. 
However, $tA1$ maps to $\epsilon$.
For this example, new blank nodes need to be created for
$tA1$ if we want our construct query to be able to transfer the values in the source $sO1_{address}$ attribute to the target  $tA1_{address\_line}$ attribute. These blank nodes correspond to existential variables that are common in data exchange~\cite{fagin2005data}.   
In the same example, the renaming
$\Re_2$ 
maps some of the other target variables such as $tO2$ to $\epsilon$.
In this case, no value invention is need to maintain the structure of the translated data, we can simply not map any data to this node.
For target attribute values, generally value invention is not required.  

To summarize, 
\methodName will create blank nodes for a variable in the target construct query if that variable is representing a concept which is not an aligned concept or if it represents an aligned concept which  
is mapped to an $\epsilon$
\emph{and} if these variables (whether they are aligned or not) are in a path that leads to mapped concepts or mapped attribute variables. 
To generate the \texttt{Construct} clause of the query, the association query pattern of  
$T$ will be used (without the \texttt{OPTIONAL} keywords). In addition, as described above, the variable names in this pattern will be converted to blank nodes, 
if necessary.  
\TRep{In practice, the above approach for creating blank nodes is inefficient since it creates redundancy. Section~\ref{bn} explains in more detail how this query can be optimized to generate the blank nodes with less redundancy.} 

\subsection{Discussion}\label{sec:discussion}
in what follows we aim to elaborate more on some of the issues that may arise during the process of mapping rule generation.

\subsubsection{Dealing with instances which have more than one most-specific types}\label{multiInherit}

In this work, we assumed that each individual has exactly  one most-specific type. According to the W3C recommendation document on OWL 2, this is a common assumption in many semantic web applications, since by assuming otherwise ``many potentially useful consequences can get lost"~\cite{owl2pri}. However, we acknowledge that a KMGT needs to support mapping generation for such KBs~\cite{gh2019} since
large KBs in particular may violate this assumption.  To extend our approach, our definition of semantic association must be extended to include $\texttt{rdfs:subClassOf}^\wedge$ paths. 

The reason for this assumption was that if we didn't assume that the most-specific types are disjoint, in our discovery phase we had to also follow $\texttt{rdfs:subClassOf}^\wedge$, since this new assumption implies that the relatedness of aligned concepts are no longer just expressed explicitly through data-level object properties of the KB (data-level to indicate that we are not talking about schema level object properties such as rdfs ones). To give an example, suppose that in a KB, instance $i$ is both an \texttt{Employee} and a \texttt{Student}, but there is no path between \texttt{Employee} and a \texttt{Student} (except of course a path in which you have to traverse to their most specific common parent from one and then from that parent traverse down to the other - note that such path exists in well formed KBs for any two concepts).  This means that \texttt{Employee} and a \texttt{Student} are associated because they are not disjoint (instance $i$ is in both). This association is not defined through explicit data-level object properties. If we want to show this type of association we have to also follow $\texttt{rdfs:subClassOf}^\wedge$ which will significantly increase the number of paths created, many of them unwanted.

Currently, to handle such cases, we go through a preparation phase, in which we create a dictionary in which each aligned concept is paired with a set of other aligned concepts which have intersection with it. This dictionary can be populated using instances of KBs and/or axioms such as \texttt{owl:disjointWith}. Note that when instances are used, one can define a threshold, expressing for instance, if there is at least 20\% intersection, assume these concepts are associated. Using this dictionary,  when we are creating associations, for each concept which is added to the association, we automatically add its paired concepts indicated in the hash map as well. Performance-wise, in real-world scenarios, this should not pose a problem, since although expanding with new concepts will increase the number of interpretations by increasing the number of object properties that needs to be traversed, we have already stressed test \methodName using very large settings. However, note that using this approach, we no longer guarantee to find all possible associations (unless of course a KB is annotated completely with \texttt{owl:disjointWith} axioms). In addition, if there is no cue in the KB that could be used to create the hash, this approach will not work at all.  It is worth mentioning that in future, it is interesting to see whether schema summarization methods such as~\cite{pham2015deriving} can help in combining concepts and pruning properties so that this adjustment is not necessary.

\subsubsection{Dealing with unnecessary blank nodes}\label{bn}
It is not hard to observe that the query in Figure~\ref{mapping} will produce unnecessary blank nodes. That is, even if none of the attributes of the blank node is bounded, since we always add a triple pattern that defines the type of the blank node, the blank node will be created. One approach to address this problem is to create and bound the blank nodes in the optional phrases of the where clause of the query (instead of defining the blank nodes in the construct clause of the query.) To demonstrate, assume that we have the following construct query:

\begin{linenumbers}
{\obeylines\obeyspaces\scriptsize
\texttt{\underline{construct}\{
~~~~-:trgt a trgt:Atype.
~~~~-:trgt trgt:AnAttribute ?trgtAttr\}
~~\underline{where}\{
~~~~?src a src:AnotherType.
~~~~OPTIONAL\{ ?src src:otherAttr ?srcAttr.\}
~~~~bind(?srcAttr as ?trgtAttr)\}
   \resetlinenumber
}} 
 \end{linenumbers}
\medskip

In this query, even if \texttt{?srcAttr} (and as a result \texttt{?trgtAttr}) is not bounded, line 2 will cause a blank node of type \texttt{trgt:Atype} to be created. This is not desirable when creating mapping rules since we only need to create a blank node if an actual data needs to be translated. In other words, we only want to create the blank node for solutions in which the optional phrase in the source is matched. To achieve this, we need to create the blank node within the optional phrases' scope. For instance, we can use the following query instead of query above.

\begin{linenumbers}
{\obeylines\obeyspaces\scriptsize
\texttt{\underline{construct}\{
~~~~?trgt a trgt:Atype.
~~~~?trgt trgt:AnAttribute ?trgtAttr\}
~~\underline{where}\{
~~~~?src a src:AnotherType.
~~~~OPTIONAL\{ ?src src:otherAttr ?srcAttr.
~~~~~~~~~bind(BNODE("trgt") as ?trgt)\}
~~~~bind(?srcAttr as ?trgtAttr)\}
    \resetlinenumber
}} \end{linenumbers}
\medskip 

Note that when using this strategy, a renaming function is defined as $\Re: T_{vars} \rightarrow S_{vars} \cup \{\epsilon\} \cup \beta$ where $\beta$ is a set of blank node IRIs.

\subsubsection{Dealing with large KBs}\label{scale}
When dealing with large general KBs, usually only a small fragment of an underlying KB is useful to the exchange problem. Trying to find associations among all elements of a very large KB can be a waste of resources. Thus we have adopted the knowledge fragment selection approach~\cite{hellmann2009learning}, to gather knowledge about the resources most important to the exchange problem at hand. Through this process, we can create a small fragment of the general KB and find associations only within this fragment. The goal of the knowledge fragment selection in general is to select relevant knowledge by traversing the knowledge graph starting from important resources up to a certain given recursion depth. This method is usually used in machine learning approaches that relies on a significant amount of reasoning. In these approaches usually a set of individuals is given as positive / negative examples, and thus the fragmentation algorithm starts from those given instances and explore the neighbours of those instances up to a given recursion depth. 

\begin{figure}
\centering
\includegraphics[width=0.5\textwidth]{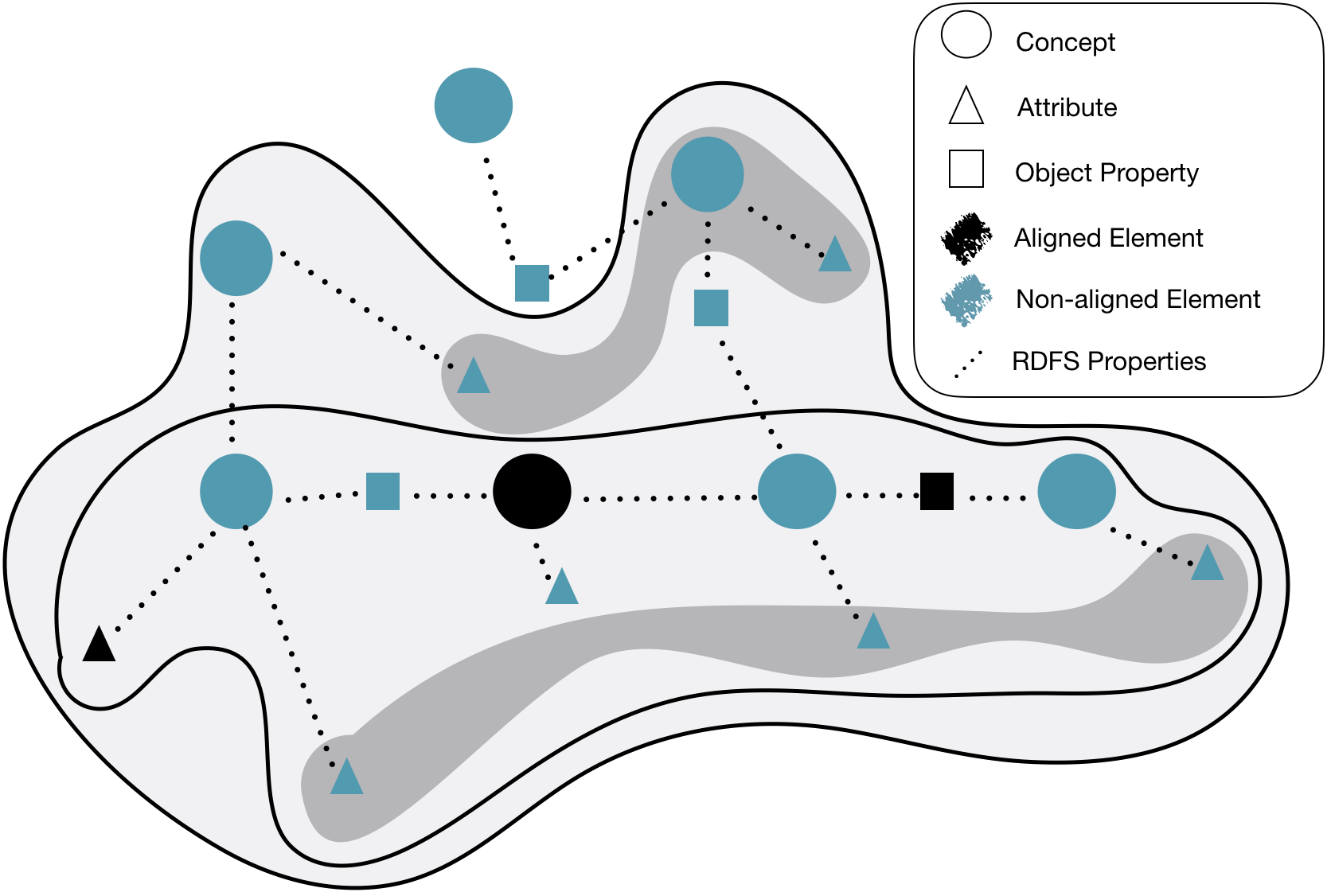}
\caption{\label{fig:expansionStep} The solid inner border represents the graph that is produce by expanding the corresponded nodes with recursion depth 1. The solid outer border represents the recursion depth 2. The darker section within the solid borders represents nodes which are not in the recursion depth i, but added to this depth since they are attributes of some nodes in this recursion depth or the object of the object property in this depth.}
\end{figure}

In our problem, the starting point of the algorithm is a set of aligned elements of the KB. If a concept is reachable from a node within a certain recursion depth, we assume that all of the concept's attributes (but not its object properties) and their values are reachable within that range. If an object property node is reachable within a certain recursion depth, we assume that both its subject and object are reachable within that depth (see Figure~\ref{fig:expansionStep}). Note that the statements which aim to identify same concepts or properties (i.e. \texttt{owl:equivalentClass, owl:equivalentProperty}) are not adding to the recursion depth. Obviously the recursion depth can affect the quality of the generated mappings. 

This is a preliminary solution. We plan to investigate whether
sophisticated methods such as modularization ~\cite{grau2008modular,konev2013model,stuckenschmidt2009modular,Ghazvinian:2011:MMU:1999676.1999684,Grau:2007:JRA:1242572.1242669} or partitioning~\cite{stuckenschmidt2004structure,grau2005automatic,Seidenberg:2006:WOS:1135777.1135785} can help in dealing with knowledge translation between large KBs.

\section{Mapping Ranking} \label{sec:ranking}

\newcommand{\dispositional}{\mbox{Path}\xspace}
\newcommand{\reachability}{\mbox{Consistency}\xspace}
\newcommand{\bounding}{\mbox{Coverage}\xspace}

For a given skeleton, the set of possible renamings, even valid renamings, each of which creates a different mapping,
might be large due to the diverse ways in which knowledge can be represented in KBs.   
Thus, we introduce a set of heuristics that help rank 
renamings of each skeleton.  
A data engineer can then browse through the ranked set of mappings (or through a set of examples created using mapping queries)
and choose the most desirable set.

Many cues in the KB can be used to rank our mappings. For instance, Maponto~\cite{an2006discovering}, which produces mapping rules in settings that involve relational models and KBs (see Section~\ref{sec:rw} for details) suggests that reach axioms such as functionality or cardinality can be used to select  better mappings. However, these reach axioms are usually not available~\cite{buhmann2012universal}. In addition, instances of source and target KBs can be used to select  mappings which exchange more facts 
that are already present in the target~\cite{Qian:2012:SSM:2213836.2213846}. However, one of the most important applications of this work is to populate a target, and thus it is not reasonable to assume that target is already populated with a large 
amount 
of the source's data. Furthermore, 
in the presence of set of  positive or negative facts,  rule mining approaches (s.a. ~\cite{galarraga2015fast,meng2015discovering,ortona2018robust}) can help identify important paths (or rules) that best fit a set of given examples.
However, we cannot always  assume that such examples exists and creating a good set of examples is itself an interesting research challenge.  
In this work, we present three complementary heuristics that do not require the information mentioned above, but that can be used in concert with other information when available. 
Our first two heuristics are structural, giving higher ranks to mappings that are more consistent with the structure of the KBs, the third is based on the coverage of the mappings, giving higher ranks to renamings that cover more target elements.

First, we rank mappings based on the degree to which source and target {\bf paths} are collectively mapped.
Recall that an association path does not go through any other aligned concept. 
If a mapping maps two target concepts between which there is an association path, to two source concepts that have no association path between them, we  
rank this mapping lower than a mapping that uses 
two source concepts which are connected with an association path.
\TRep{
\begin{example}
In Figure~\ref{fig:saEmp}, \texttt{sC3} represents a \texttt{Country} (the country in which an employee was born) and 
some mappings will map it to \texttt{tC1}.
However, such mappings are less desirable than 
mappings that map either \texttt{sC1} or \texttt{sC2} to \texttt{tC1}.
To see why, note that \texttt{sC3} represents countries which are directly related to employees.
While, in the 
target KB there is no such direct association between employees and countries. 
On the other hand,
\texttt{sC1} or \texttt{sC2} represent \texttt{Countries} which are related to \texttt{Employee} via \texttt{src:Organization} (which is mapped to \texttt{trgt:Organization}), and there are
association paths between (\texttt{trgt:Employee},\texttt{trgt:Organization}) and (\texttt{trgt:Organization},\texttt{trgt:Country}) in the target. In other words, the target KB 
models countries as locations of organizations 
so mappings that use source countries that are also locations of organizations are likely to be better models of the domain.
\qed
\end{example}
Our {\bf \dispositional Priority} ranking is based on this intuition.}
\theoremstyle{definition}
\begin{definition} {\bf (\dispositional Priority)} 
Assume $s_0 \leadsto_{c2c} t_0$. For any association path $p$ with root $s_0$, the cost of adding $p$ to $s_0$  with respect to the $target$ KB, $Cost(s_0,p)$, is 1 if there exists no association path between $t_0$ and $t_1$, where $s_1 \leadsto_{c2c} t_1$, 
and $s_1$ is the tail of $p$, and is 0 otherwise.

The cost of adding a set of ordered association paths, $ \{p_1,...p_k\}$, to $s_0$, where $k > 2$, and $s_0$ is the root of $p_1$, $s_i$ is the tail of $p_i$ and the root of $p_{i+1}$,  
with respect to the $target$ KB, $Cost(s_0,p_k)$, is:
\[ Cost(s_0,p_{1}) + \sum_{i=1}^{k-1} Cost(s_{i}, p_{i+1})\]

\noindent For node $n$ in $\sa{u_0}$, where $concept(n)$ is an aligned concept, a set of ordered association paths, $P$, can be defined such that $P$ includes association paths that are used to connect $n_0$ to $n$, in that order, where $n_0$ is the root of the association tree. The \dispositional score of variable $v$ of node $n$ is the cost of adding $P$ to $u_0$, and is  $score(v, \sa{u_0})$. 
For Skeleton $\langle S,T \rangle$, The \dispositional cost of $\Re$ is: 
\[Path_{cost}(\Re,\langle S,T \rangle) = \sum_{v} score(v,S)\] 
where $v$ ranges over all the variables of nodes of aligned concepts in $S$.
We say $\Re_{i}$ has higher \dispositional priority than $\Re_{j}$ 
if it has a lower cost. \qed
\end{definition}

\begin{example}\label{shortestPath}
In Figure~\ref{fig:saEmp}, the $score(\texttt{sC3},\sa{\texttt{src:Employee}})$ is 1, since there is no association path between \texttt{trgt:Employee} and \texttt{trgt:Country} in the \textbf{target} KB, while $score(\texttt{sC1},\sa{\texttt{src:Employee}})$ or $score(\texttt{sC2},\sa{\texttt{src:Employee}})$ are 0. Note that sometimes it is tempting to pick the shortest path between two concepts as the best property path that describes the relationship between them 
(e.g., \texttt{src:born\_in} here), however this example shows that the shortest path is not always best. \qed
\end{example}

\TRep{
The \dispositional heuristic 
ranks higher mappings that faithfully map paths of aligned concepts. 
 Of course, there are times where this heuristic 
is incorrect.  For instance, assume that in the source KB of Figure~\ref{mainex}, instead of the attribute \texttt{src:address} we had an actual concept \texttt{src:Address} which was connected to \texttt{src:Organization} via \texttt{src:hasAddress} object property. Also, assume that $\texttt{src:Address} \leadsto_{c2c} \texttt{trgt:Address}$. Finally, assume that \texttt{src:Person} is connected to \texttt{src:Address} via \texttt{src:hasHomeAddress}. In this case, the renaming that picks the home address of a person as an organization's address will get higher priority. The reason is that, in the target, address is the property of a person and not the organization, and thus the address which directly describes a person in the source, \texttt{src:hasHomeAddress}, (and not the organization) will be picked. 
}

Next, we consider semantic associations that include multiple subpaths containing identical concepts (like the two \texttt{Organization, Country} paths in our running example).  These paths will have identical \dispositional scores.  Mappings that only use a single one of these multiple paths (rather than mixing concepts in two or more such identical paths) are generally better reflections of the domain.
\TRep{
\begin{example}\label{reachabilityP}
In Figure~\ref{fig:saEmp}, \texttt{sC1} and \texttt{sC2} both represent the source concept \texttt{Country}. 
Assume there are renamings $\Re_y$ such that $\Re_y\texttt{(tO2)} = \texttt{sO2} $, and $\Re_y\texttt{(tC2)} = \texttt{sC1}$ and $\Re_z$ such that $\Re_z\texttt{(tO2)
}= \texttt{sO2}$, and $\Re_z\texttt{(tC2)} = \texttt{sC2}$. 
It is not hard to see that $\Re_z$ is 
almost certainly  
more desirable. Intuitively, 
it does not make sense to transfer the country of an employee's former \texttt{Organization} (\texttt{has\_worked\_for}) to target and associate it with the country of her current \texttt{employer}, or vice versa.  One would almost always want an organization to be translated with its own country.
Our {\bf \reachability Priority} aims to capture 
this intuition.\qed
\end{example}
}

To define {\bf \reachability Priority}, we note that the variables in a semantic association (either $S$ or $T$) are organized in a tree starting at the variable for the root.  We say $v_j$ is reachable from $v_i$, if $v_j$ 
is a descendant of $v_i$ in this tree.\\  
\theoremstyle{definition}
\begin{definition}{\bf (\reachability Priority)} 
For skeleton  
$\langle S,T \rangle$, set the \reachability cost of the renaming $\Re$, $Consistency_{cost}(\Re,\langle S,T \rangle)$, to zero.
For each pair of target variables $v_i$ and $v_j$ which both are bounded by some source variables in $\Re$:
\squishlist
\item
if $v_j$ is reachable from $v_i$ (in $T$),  
and $\Re(v_j)$ is {\bf not} reachable from $\Re(v_i)$ (in $S$), then increase 
$Consistency_{cost}(\Re,\langle S,T \rangle)$ by 1; 
\item
if $v_j$ is {\bf not} reachable from $v_i$ (in $T$) and $\Re(v_j)$ {\bf is} reachable from $\Re(v_i)$ (in $S$), then increase  
$Consistency_{cost}(\Re,\langle S,T \rangle)$ by 1;
\squishend
We say $\Re$ has higher priority if it has lower cost.
\qed
\end{definition}
\TRep{
\reachability priority might wrongly prioritize renaming functions. One obvious case is when two mapped concepts which are associated using a property path $p$ in the source are associated with $p^{\wedge}$ in the target. \methodName
takes this into account when the mapping is from a property to a reverse property.}
 
In addition to the two structural heuristics, we also use 
{\bf \bounding Priority} which prioritizes renamings that contain translations for (or cover) more target elements.  
\theoremstyle{definition}
\begin{definition}{\bf (\bounding Priority)} 
For a skeleton $\langle S,T \rangle$, \bounding cost of $\Re$, $Coverage_{cost}(\Re,\langle S,T \rangle)$, is the number of variable in T that are  
mapped to $\epsilon$. We say $\Re$ has higher priority if it has lower cost.
\qed
\end{definition}

\section{Evaluation}\label{eval}
We begin by 
discussing current benchmarks for knowledge exchange and 
comparing \methodName with other KGMTs in handling the scenarios in these benchmarks.
Then, in Section~\ref{usecase}, we showcase some of the most important real-world applications of KB translation and show the effectiveness of \methodName in these contexts.
In Section~\ref{synth}, we use 50 synthesized settings to stress test the performance of \methodName.
Note that we use a version of Kensho which creates renamings in which each source variable is assigned to at most one target variable. This strategy significantly reduces the number of possible renamings produced, at a cost of not being able to produce some valid renamings in rare cases.
We have also done a small case study which compared the results of data engineers manually writing mapping rules vs.~selecting, using data examples, mappings generated by \methodName.

\subsection{Benchmark Evaluation}\label{benchmark}
STBenchmark~\cite{alexe2008stbenchmark} introduced the use of (micro) scenarios for comparing data exchange systems that use a structured or semi-structured model.  This idea was generalized by 
the meta-data generator iBench~\cite{arocena2015ibench} that permits the efficient creation of benchmarks with large and complex schemas and data exchange scenarios.
DTSBenchmark~\cite{rivero2011benchmarking}  
provides a set of scenarios where the source and target are both KBs. These scenarios were later refined in LODIB (linked open data integration benchmark)~\cite{rivero2012benchmarking}
which is mainly designed to benchmark the expressive power of mapping languages. Both Mosto and \methodName can automatically generate the desirable mapping rules for all the scenarios proposed in DTSBenchmark. In addition, it is reported that queries generated by Mosto can support the expression of all fifteen LODIB scenarios except for three (the ones which need conditional clauses or aggregation queries)~\cite{rivero2012benchmarking}. Queries generated by \methodName have the same expressive power. 
\methodName, like Mosto, cannot automatically learn the relationship between values (e.g., when a value must be transformed using function \texttt{usDollarsToEuros}), and thus it cannot automatically generate the mapping rules for the LODIB scenarios that require these types of transformations.

Additional scenarios have recently been identified that need to be supported by KMGTs~\cite{gh2019}.  Two of these scenarios are inspired by value invention in relational data exchange.  A third involves handling source KBs that are incomplete.  A fourth involves mapping creation even when the set of given correspondences are incomplete (a very common case in practice).  Finally, a fifth scenario involves mappings that use cyclic property paths.  \methodName can handle all five scenarios while the previous approaches (Mosto and Qin et al.) cannot.
It is important to note that since the associations that existing KMGTs create are a subset of what Kensho creates, and since their mapping language is not more expressive than Kensho's, there would be no scenario in which these KMGTs can create a mapping which Kensho cannot.

\subsection{Knowledge Translation Usecases}\label{usecase}

\begin{table*}[]
\caption{
Mapping Generation and Ranking Performance. 
Generation time is not mentioned if it is $<1$ sec.}
\resizebox{\textwidth}{!}{%
\begin{tabular}{|c|c|c|c|c|c|c|c|c|c|c|c|c|}
\hline
\rowcolor[HTML]{B3BDDD} 
\cellcolor[HTML]{B3BDDD}                               & \cellcolor[HTML]{B3BDDD}                                                                           & \cellcolor[HTML]{B3BDDD}                                                                      & \cellcolor[HTML]{B3BDDD}                                                       & \multicolumn{4}{c|}{\cellcolor[HTML]{B3BDDD}\textbf{Number of mapping generated}}                                                                                                                                                                                                                                                                                                              & \multicolumn{4}{c|}{\cellcolor[HTML]{B3BDDD}\begin{tabular}[c]{@{}c@{}}\textbf{Effect of Ranking}\\ $\Re_{gold}$'s rank--\# $\Re$s ranked better--\# $\Re$s ranked equla\end{tabular}} & \cellcolor[HTML]{B3BDDD}                                      \\ \cline{5-12}
\rowcolor[HTML]{B3BDDD} 
\multirow{-2}{*}{\cellcolor[HTML]{B3BDDD}\textbf{Sec}} & \multirow{-2}{*}{\cellcolor[HTML]{B3BDDD}\textbf{Source}}                                          & \multirow{-2}{*}{\cellcolor[HTML]{B3BDDD}\textbf{Target}}                                     & \multirow{-2}{*}{\cellcolor[HTML]{B3BDDD}\textbf{Corr}}             & \textbf{\basic}                                                                                         & \textbf{\rel}                                                                                 & \textbf{\atr}                                                                          & \textbf{\methodName}                                                                         & \textbf{\bounding}                                   & \textbf{\dispositional}                                   & \textbf{\reachability}                                  & \textbf{\dispositional + \reachability}                                  & \multirow{-2}{*}{\cellcolor[HTML]{B3BDDD}\textbf{Similarity}} \\ \hline
\cellcolor[HTML]{B4D9B9}\ref{patent}                   & \begin{tabular}[c]{@{}c@{}}\#classes(C): 64\\ \#Attributes(A):67\\ \#objectProp(P):69\end{tabular} & \begin{tabular}[c]{@{}c@{}}C: 10\\ A:2\\ P:11\end{tabular}  & \begin{tabular}[c]{@{}c@{}}\#c2c: 2\\ \#atr2atr: 1\\ \#rel2rel: 0\end{tabular} & \begin{tabular}[c]{@{}c@{}} 2\end{tabular}                      & \begin{tabular}[c]{@{}c@{}} 2\end{tabular}         & \begin{tabular}[c]{@{}c@{}} 2\end{tabular}    & \begin{tabular}[c]{@{}c@{}} 2\end{tabular}   & 1-0-0                                               & 2-1-0                                                    & 1-0-1                                                  & 2-1-0                                                  & \begin{tabular}[c]{@{}c@{}}Low \\ (2.6\%)\end{tabular}         \\ \hline
\cellcolor[HTML]{B4D9B9}\ref{migration}                & \begin{tabular}[c]{@{}c@{}}C: 5\\ A:2\\ P:2\end{tabular}         & \begin{tabular}[c]{@{}c@{}}C: 6\\ A:3\\ P:3\end{tabular}    & \begin{tabular}[c]{@{}c@{}}\#c2c: 5\\ \#atr2atr: 2\\ \#rel2rel: 2\end{tabular} & \begin{tabular}[c]{@{}c@{}} 7\end{tabular}                      & \begin{tabular}[c]{@{}c@{}} 4\end{tabular}         & \begin{tabular}[c]{@{}c@{}} 7\end{tabular}     & \begin{tabular}[c]{@{}c@{}} 4\end{tabular}   & 1-0-0                                               & 1-0-3                                                    & 1-0-3                                                  & 1-0-3                                                  & \begin{tabular}[c]{@{}c@{}}High\\ (91\%)\end{tabular}         \\ \hline
\cellcolor[HTML]{B4D9B9}\ref{alignment}                & \begin{tabular}[c]{@{}c@{}}C: 38\\ A: 23\\ P: 13\end{tabular}    & \begin{tabular}[c]{@{}c@{}}C: 49\\ A:11\\ P:17\end{tabular}  & \begin{tabular}[c]{@{}c@{}}\#c2c: 4\\ \#atr2atr: 1\\ \#rel2rel: 0\end{tabular} & \begin{tabular}[c]{@{}c@{}} 26\end{tabular}                     & \begin{tabular}[c]{@{}c@{}} 26\end{tabular}         & \begin{tabular}[c]{@{}c@{}} 26\end{tabular}    & \begin{tabular}[c]{@{}c@{}} 26\end{tabular}   & 2-6-11                                              & 1-0-25                                                   & 3-15-3                                                 & 3-15-3                                                 & \begin{tabular}[c]{@{}c@{}}Low\\ (10\%)\end{tabular}          \\ \hline
\cellcolor[HTML]{B4D9B9}\ref{synthSecen}               & \begin{tabular}[c]{@{}c@{}}C: 6\\ A: 12\\ P: 5\end{tabular}      & \begin{tabular}[c]{@{}c@{}}C: 5\\ A: 14\\ P: 6\end{tabular} & \begin{tabular}[c]{@{}c@{}}\#c2c: 4\\ \#atr2atr: 7\\ \#rel2rel: 1\end{tabular} & \begin{tabular}[c]{@{}c@{}} 407 $\times 10^5$\\ creation time:\\ $\sim 2.5$ hr\end{tabular} & \begin{tabular}[c]{@{}c@{}} 65700\\ creation time:\\ $\sim 10$ min\end{tabular} & \begin{tabular}[c]{@{}c@{}} 4015\end{tabular} & \begin{tabular}[c]{@{}c@{}} 657\end{tabular} & 1-0-447                                             & 1-0-20                                                   & 1-0-15                                                 & 1-0-3                                                  & \begin{tabular}[c]{@{}c@{}}High\\ (81\%)\end{tabular}         \\ \hline
\end{tabular}%
}
\label{comp}
\end{table*}

We now investigate the effectiveness of  
mapping generation and ranking strategies implemented in \methodName using several scenarios that 
showcase some of the important applications of KB translation. 
Note that it is important to evaluate Kensho on real scenarios because all mappings which are created by our algorithm are valid interpretations of the set of correspondences, however, depending on the context, some mappings may be more desirable than others.
In all settings presented, 
an expert identified
the desired (gold) mapping rule in the selected skeleton, henceforth called $\Re_{gold}$. 
For each scenario (described below), in
Table~\ref{comp},  
column {\basic} reports the number of mappings which are {\basic} valid, column {\atr} reports the number of mappings which are {\basic} and \emph{\atr} valid, column {\rel} reports the number of mappings which are {\basic} and \emph{\rel} valid , and column \methodName shows number of mappings that adhere to all of our validity constraints.
Table~\ref{comp} also shows the effect of our ranking methods in facilitating the selection of the best set of mappings.
For the different ranking strategies, we report on 
the rank of $\Re_{gold}$ in the \methodName rankings, and since our rankings can have ties, we report on the number of other mappings ranked better than $\Re_{gold}$, then the number of mappings ranked equal to $\Re_{gold}$ (cols.~9-12).
For example, 2-1-0 means $\Re_{gold}$ was ranked second, with only one other mapping ranked ahead of $\Re_{gold}$ and with no ties.
The last column of Table~\ref{comp} reports the similarity between the two KBs (informally, the number of resources that model the same real world entities).

\subsubsection{Populating a KB using Open Data}\label{patent}

We believe the most important application of Kensho is populating an existing domain-specific ontology using other currently available structured data sources where the data source does not have to be a KB as long as it can be automatically converted to one. 
Expertise finding is one area for which having a domain-specific KB can be very beneficial~\cite{hansen2004build,paquette2007ontology}. 
For this reason, there are carefully designed ontologies in the literature that model expertise for competency management.
In this setting, for our target KB, we use  
one of these ontologies~\cite{fazel2012ontology} and augment it with
two more concepts, \texttt{Employee} and \texttt{Organization}, to be able to  
model the skills of employees in various companies. Our goal in this scenario (Row 1 of Table~\ref{comp}) is to generate mapping rules that can be used to populate this KB using open data published by the US Patent and Trademark Office (USPTO).
To use the USPTO corpus as our source KB, we started with a subset of the USPTO's patent XML corpus~\cite{pat} 
and automatically 
created a linked data corpus from it 
using Xcurator~\cite{yeganeh2011linking} and further enriched it using Vizcurator~\cite{GhadiriBashardoost:2015:VVT:2740908.2742845}. 
The original set of correspondences was provided by a domain expert and then was verified by the crowd using 
the protocol proposed by Sarasua et al.~\cite{sarasua2012crowdmap}. 

Our two KBs mainly model different domains (one models skills in an organization and the other models patent applications), so the similarity between these two KB
is low with only 2.6\% of the concepts and properties modeling the same information (last column of Table~\ref{comp}).
Additionally, there are two correspondences between concepts and only one between attributes (col.~4).
Mosto and Qin et al.~\cite{qin2007discovering} cannot find the gold mapping which connects the associated concepts and 
attributes through properties with no correspondences.
\methodName produces two possible  mappings  
including the gold mapping.  

One side benefit of our approach is that the results can be used to enrich the 
alignment. For instance, in this scenario $\Re_{gold}$ suggests that a 
Rel2Rel correspondence may exists between the source \texttt{Patent - Inventor} path and the target \texttt{hasSkill - Skills} path.

\subsubsection{Migrating Data between KBs}\label{migration}
This 
example (Row 2 in Table~\ref{comp}) is 
from Mosto~\cite{rivero2013exchanging}, the source is a portion of DBpedia version 3.2 and the target is the similar portion of DBpedia version 3.6.
In order to generate the mapping rules in this scenario, Mosto 
requires the correspondences and also 
a user-provided
 axioms called \texttt{mosto:strongRange}. 
 Note that \methodName does not require users to specify axioms such as these to guide the algorithm. 
In Mosto, if two source concepts ($s_1$ and $s_2$) have correspondences to two target concepts ($t_1$ and $t_2$) then these 
concepts will only be included in the same mapping if both the source and the target concepts are in a subclass relation, or if they are the domain and range of an \emph{aligned} property, or if a data engineer has manually added a \texttt{mosto:strongRange}
or \texttt{mosto:strongDomain} relation between them.
In this scenario, without the annotation, Mosto would exchange the \texttt{Actor} and \texttt{AcademyAward} data in DBpedia, but not the relationship between these two concepts (representing about 850 facts indicating who won which award).

In this scenario, 
\methodName generates four mappings.  If we do not use our r2r strategy, then we generate seven mappings.
(including some that are inconsistent with the Rel2Rel correspondences).  The four mappings that \methodName produces include $\Re_{gold}$ and another three that are less complete (containing some existential target variables).  Our {\bf \bounding} ranking strategy
correctly ranks the gold mapping before these more incomplete mappings.  

\subsubsection{Enriching the Result of Ontology Alignment}\label{alignment}
In this  
scenario (Row 3),  
we highlight 
the fact that  
\methodName can enrich the rules produced by the current state-of-the-art alignment tools. Note that while alignment tools generate correspondences, \methodName enriches these correspondences and produces \emph{executable queries } 
by interpreting the  
correspondences collectively. 
Previous to this work, KMGTs assumed that the set of property correspondences are complete. Thus, if there is no correspondence
to or from an object property, the KMGT did not 
automatically consider any path that included that property.
One of the main advantages of \methodName
is that it 
considers such properties in mapping generation.
To show the benefit of this approach,
in this experiment, we have used two KBs from the OAEI (Ontology Alignment Evaluation Initiative) campaign and the correspondences between them which were produced by AML (Agreement Maker Light) -- one of the best performers in the 2018 matching challenge~\cite{faria2018results}.
The KBs we have used, \emph{confTool} and \emph{Sigkdd}, are from the OntoFarm dataset~\cite{zamazal2017ten}  
of the {\it Conference} track of the OAEI.

In this scenario, \methodName produces 26 mappings (including the gold mapping).
These two KBs are not very similar, so our structural rankings are not helpful.  For example, our {\bf \dispositional} ranking ranked all 26 mappings equally. Our
{\bf Consistency} 
ranking did better with only  
eighteen mappings ranked equal or higher than the gold mapping. Our results suggest some interesting possible correspondences between two KBs which are not provided in the results of AML, for instance, inverse property correspondences such as: 

$\texttt{sigkdd:submit}^\wedge \leadsto_{r2r}
\texttt{conftool:writtenBy}$\\
In this scenario, the number of 
given property correspondences
is very low. As a result, existing KMGTs 
do not create the gold mapping.

\subsubsection{Translating KBs with Cycles}\label{synthSecen}
In this scenario (Row 4),
we show that unlike existing KMGTs,
\methodName can handle cycles, a situation that occurs in many existing KBs such as those created from social networks. Cycles define associations between resources of the same type such that multiple resources in the source can be mapped into multiple resources in the target, a situation resulting in a large number of possible mappings. This scenario showcases that our mapping generation and ranking strategies are especially useful in cases where the number of possible interpretations becomes very large due to the existence of cyclic paths. 

This setting (depicted in Figure~\ref{mainex}) was created 
from the example proposed by Qin et al.~\cite{qin2007discovering} which was developed using Carnegie Mellon's \emph{Person \& Employee Ontology} and the University of Maryland's \emph{People Ontology}. To add complexity, we incorporated a cycle by adding an additional object property, \texttt{src:related}, which has \texttt{src:Employee} as its 
domain and range. 
This caused a large number of mappings to be created. However, our mapping generation validity strategies reduce the number of mappings to a few hundred (which is still unmanageable for a human), but the {\bf Consistency} ranking ranks the gold mapping first, in a tie with only 15 other mappings (a much more reasonable task for an engineer to understand).  Together,  our Path and Consistency ranking reduces the ties to only three.

\subsection{Performance Evaluation}\label{synth}
Mapping generation tools are usually evaluated using metadata generators~\cite{alexe2008stbenchmark,arocena2015ibench}.
They 
allow the data engineer to systematically vary specific parameters that influence the difficulty of relational mapping creation or data exchange.
In a similar vein, MostoBM~\cite{rivero2012benchmarking2}
identifies 
three schema-level 
parameters (namely, {\bf $L$ or depth} of the class relationships, {\bf $C$ or breadth} 
of the class relationships, and {\bf $D$}, the number of attributes) that can affect the complexity of the task of KB mapping generation. To investigate the effect of breadth, 
we followed the MostoBM approach and 
generated ten settings for one of 
the MostoBM exchange scenarios - called \emph{sink properties}. More specifically, we fixed $D$ = 10,
$L$ = 1, and vary the value of $C$ between one and ten. We call this group of settings that vary $C$, $E_1$.
For the first setting of $E_1$, \texttt{\{L=1, C=1, D=10\}}, the source contains two concepts, $\texttt{A}_0$ and its child $\texttt{A}_1$, and $\texttt{A}_0$ is the domain of ten attributes \{$\texttt{d}_1$,...,$\texttt{d}_{10}$\}. The target has the same structure, except that the domain of $\texttt{d}_1$ is $\texttt{A}_1$. Resources with the same labels correspond to each other (e.g., $\texttt{src:A}_0 \leadsto_{c2c} \texttt{trgt:A}_0$). For the next setting in $E_1$, \texttt{\{L=1, C=2, D=10\}}, $\texttt{A}_0$ has one more child, $\texttt{A}_2$, in both the source and target and the domain of $\texttt{d}_2$ in the $target$ is $\texttt{A}_2$.
We repeated the same procedure for investigating the effect of depth and created a group of ten depth settings, $E_2$
by keeping $C$ constant and varying $L$. The difference here is that the new concept for each consecutive setting will be nested inside the most specific type (as opposed to being added to the root). For instance, 
when $L=2$
, $\texttt{A}_2$ is a child of $\texttt{A}_1$ and $\texttt{A}_1$ is a child of $\texttt{A}_0$. In this section, we report on the number of $all$ mappings created (as opposed to the number of mappings created for a specific skeleton as was done in the previous section). Note that changing $D$ while fixing $L$ and $C$ will not change the number of mappings created and so we did not include such settings in our evaluation.

\begin{figure}\small
  \centering
  \vspace{0pt}
  \begin{minipage}[b]{0.2\textwidth}
  \begin{tikzpicture}
	\begin{axis}[
		xlabel=\scriptsize{x},
		ylabel=\scriptsize{\# of mappings},
		legend columns=-1,
        legend style ={ at={(0.1,1.1)}, 
        anchor=south west, 
        },
        cycle list name=exotic,
        width = 4.5cm
		]

\addplot [smooth,red,thick]  coordinates{

( 1 , 3)
( 2 , 5)
( 3 , 7)
( 4 , 9)
(5,11)
(6,13)
(7,15)
(8,17)
(9,19)
(10,21)

};
\addlegendentry{$E_1$\texttt{=\{L=1, C=x, D=10\}}};

\addplot [smooth,brown,thick]  coordinates{

( 1 , 3)
( 2 , 6)
(3,10)
(4,15)
(5,21)
(6,28)
(7,36)
(8,45)
(9,55)
(10,66)

};
\addlegendentry{$E_2$\texttt{=\{L=x, C=1, D=10\}}};

\end{axis}
\end{tikzpicture}

  \end{minipage}
  \hspace{0.2cm}
  \begin{minipage}[b]{0.2\textwidth}
      \begin{tikzpicture}
	\begin{axis}[
		xlabel=\scriptsize{x},
		ylabel=\scriptsize{time (sec)},
        legend style ={ at={(0,1)}, 
        anchor=north west, 
        font=\tiny,
        },
        cycle list name=exotic,
        width = 4.5cm
		]

\addplot [smooth,red,thick]  coordinates{

( 1 , 0.347)
( 2 , 0.438)
( 3 , 0.521)
( 4 , 0.592)
(5, 0.695)
(6,0.808)
(7,0.918)
(8,1.021)
(9,1.098)
(10,1.251)

};

\addplot [smooth,brown,thick]  coordinates{

( 1 , 0.347)
( 2 , 0.512)
(3,0.631)
(4,0.805)
(5,1.018)
(6,1.100)
(7,1.324)
(8,1.591)
(9,1.782)
(10,2.066)

};

\end{axis}
\end{tikzpicture}
  \end{minipage}
  \vspace{-4mm}
\scriptsize  \caption{Increasing Breadth(C) $E_1$ \& Depth(L) $E_2$.}\label{DvsB}
\end{figure}
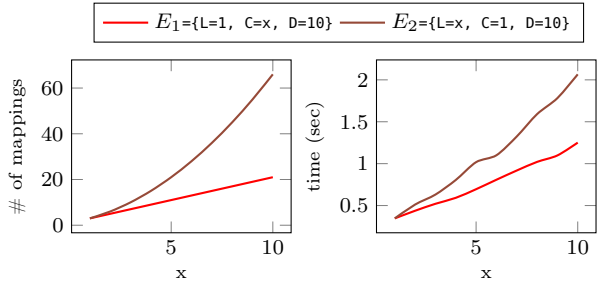

Figure~\ref{DvsB} 
shows
the number of mappings generated by \methodName in each of the settings. 
Note that Mosto does not create all possible mappings and thus in our evaluation we could not compare it with  \methodName.
As the breadth ($C$) of the KB increases, the number of mappings generated 
also increases linearly and \methodName's performance also scales linearly, remaining under 
a second 
for a 
setting with breadth of 8.  
In contrast, increasing the depth ($L$) has an exponential effect on the number of mappings generated. This is expected since increasing the depth actually adds a layer of nesting.  \methodName's performance passes one second at a depth of five which corresponds to the generation of 21 mappings. 

\begin{figure}\small
  \vspace{0pt}
  \begin{minipage}[b]{0.2\textwidth}
  \begin{tikzpicture}
	\begin{axis}[
		xlabel=\scriptsize{x},
		ylabel=\scriptsize{\# of mappings},
		ymode=log,
        log basis y={2},
        legend columns=-1,
        legend style ={ at={(-0.40,1.1)}, 
        anchor=south west, 
        },
        cycle list name=exotic,
        width = 4.5cm
		]

\addplot [smooth,red,thick]  coordinates{

( 1 , 3)
( 2 , 5)
( 3 , 7)
( 4 , 9)
(5,11)
(6,13)
(7,15)
(8,17)
(9,19)
(10,21)

};
\addlegendentry{{{\tiny$E_1$}\scriptsize\texttt{=\{L=1,C=x,D=10\}}}};

\addplot [smooth,brown,thick]  coordinates{

( 1 , 3)
( 2 , 6)
(3,10)
(4,15)
(5,21)
(6,28)
(7,36)
(8,45)
(9,55)
(10,66)

};
\addlegendentry{{{\tiny$E_2$}\scriptsize\texttt{=\{L=x,C=1,D=10\}}}};

\addplot [smooth,orange,thick]  coordinates{

( 1 , 4)
( 2 , 10)
(3,21)
(4,41)
(5,78)
(6,148)
(7,283)
(8,547)
(9,1068)
(10,2089)

};
\addlegendentry{{{\tiny$E_3$}\scriptsize\texttt{=\{L=x,C=2,D=10\}}}};

\end{axis}
\end{tikzpicture}

  \end{minipage}
  \hspace{0.45cm}
  \begin{minipage}[b]{0.2\textwidth}
      \begin{tikzpicture}
	\begin{axis}[
		xlabel=\scriptsize{x},
		ylabel=\scriptsize{$\Delta$},
        legend style ={ at={(0,1)}, 
        anchor=north west, 
        font=\tiny,
        },
        cycle list name=exotic,
        width = 4.5cm
		]

\addplot [orange,very thick, mark=triangle]  coordinates{

( 1 , 2)
( 2 , 1)
(3,1)
(4,4)
(5,8)
(6,13)
(7,19)
(8,26)
(9,34)
(10,43)

};

\addplot [smooth,red,thick]  coordinates{

( 1 , 2)
( 2 , 2)
( 3 , 2)
( 4 , 2)
(5,2)
(6,2)
(7,2)
(8,2)
(9,2)
(10,2)

};

\addplot [smooth,brown, mark = star, semithick]  coordinates{

( 1 , 2)
( 2 , 1)
(3,1)
(4,4)
(5,8)
(6,13)
(7,19)
(8,26)
(9,34)
(10,43)

};

\end{axis}
\end{tikzpicture}
  \end{minipage}
  \vspace{-4mm}
  \caption{ Increasing  Breadth(C) $E_1$  \& Depth(L)  $E_2$ and $E_3$.  $\Delta = $ \#mappings - \#correspondences }\label{DvsB3}
\end{figure}
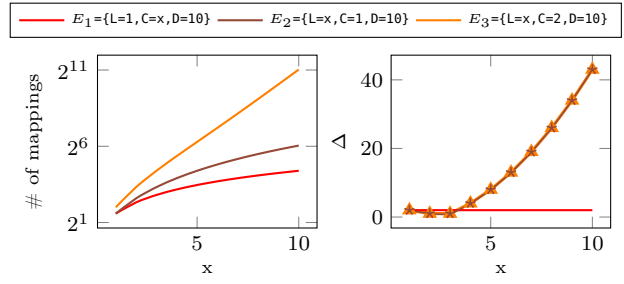

Settings in $E_1$ and $E_2$ include relatively small KBs. To push \methodName further, we  created 
a set of 10 settings, $E_3$, for the same scenario using parameters \texttt{\{L=x, C=2, D=10\}}, where $1 \leq \texttt{x} \leq 10$. The taxonomy created using these parameters contains $2^\texttt{n}$ concepts at depth \texttt{n} from the root. For instance, when \texttt{L = 10}, both the source and target contain $2^{10}$ most specific types (in total each contains more than 2000 concepts - more than three times greater than the number of concepts in DBpedia).
Figure~\ref{DvsB3} left, shows the number of mappings created in  
$E_3$ vs.~$E_1$ and $E_2$, note the scale change on the Y-axis. 
When \texttt{L = 10}, \methodName creates more than 2000 mappings and the creation time is almost three hours. One factor that affects the number of mappings is the number of correspondences. In 
$E_3$, the number of correspondences grows as the depth $L$ increases. 
To explain this effect, we have also plotted \textbf{$\Delta$} as the difference between the number  of mapping rules generated for each setting and the number of correspondences. The right plot in Figure~\ref{DvsB3} shows the result. Obviously, the difference \textbf{$\Delta$} is constant for the 
$E_1$ (breadth) settings (all have $L=2$). However it is important to note that for 
$E_2$ and $E_3$, \textbf{$\Delta$} is exactly the same for each value of the depth $L$.
This is because the number of possible mappings becomes greater than number of correspondences if there are various ways for interpreting correspondences. As we have seen, correspondences can be interpreted in various ways only if the resources can be associated with each other in multiple ways. 
In this scenario, multiple interpretations are the result of various ways that one can interpret Concept2Concept correspondences in conjunction with Attr2Attr correspondences. Thus increasing only the number of concepts and Concept2Concept correspondences (as we are doing here) will not result in multiple interpretations of the correspondences.

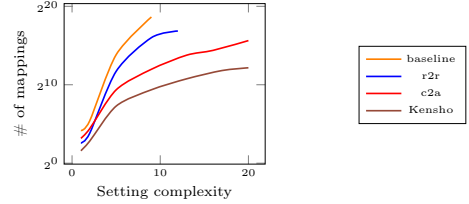
\begin{figure}\tiny

  \centering 
  \resizebox{.35\textwidth}{!}{
  \begin{tikzpicture}
	\begin{axis}[
		xlabel=\scriptsize{Setting complexity},
		ylabel=\scriptsize{\# of mappings},
        legend style ={ at={(2,0.25)}, 
        anchor=south east, 
        font=\tiny,
        },
        ymode=log,
        log basis y={2},
        cycle list name=exotic,
        width = 5cm
		]

\addplot [smooth,orange,thick]  coordinates{

( 1 , 18)
( 2 , 42)
( 5 , 14508)
( 9 , 407340)
};
\addlegendentry{baseline};

\addplot [smooth,blue,thick]  coordinates{

( 1 , 6)
( 2 , 14)
( 5 , 3348)
( 9 , 65700)
( 12 , 117936)

};
\addlegendentry{r2r};

\addplot [smooth,red,thick]  coordinates{

( 1 , 9)
( 2 , 21)
( 5 , 651)
( 9 , 4015)
( 13 , 13965)
( 16 , 21352)
( 20 , 50610)
};
\addlegendentry{c2a };

\addplot [smooth,brown,thick]  coordinates{

( 1 , 3)
( 2 , 7)
( 5 , 155)
( 9 , 657)
( 13 , 1729)
( 17 , 3587)
( 20 , 4579)
};
\addlegendentry{Kensho};

\end{axis}
\end{tikzpicture}
}

  \scriptsize
  \caption{
  Effect of increasing \# of interpretations.  
  } \label{stress}
\end{figure}

Nonetheless, \methodName is sensitive to the number of possible interpretations of a correspondence, so we created a final set of 20 settings ($E_4$) to understand how far we can push \methodName on this dimension.
We have used the setting represented in Figure~\ref{mainex}, our running example which already includes multiple interpretations, as our least complex setting (Setting 1).  To create the rest of the settings, we injected 
elements into the source, such that in each setting, the number of possible assignments for each target variable is increased by one. For instance, Figure~\ref{mapping} shows that there are three source variables of type \texttt{country} that can be assigned to variable \texttt{tC1}.  The next setting contains four possibilities for this variable, and so on. Figure~\ref{stress} represents the result of our various mapping strategies. 
The number of mappings generated grows exponentially; however, this figure demonstrates that \methodName is able to reduce this search space by only generating valid mappings.
In our largest setting, the source KB contains nearly 120 object properties and 5 concepts leading to over 500K interpretations (we have kept the target fixed in this experiment). Only one fifth of these properties corresponds to a property in the target. Note that real world KBs usually contain many fewer object properties among concepts. \methodName created 4579 possible mappings for this setting. For the biggest skeleton in this setting, \methodName ranked 42 mappings as rank one. This highlights one of the weaknesses of \methodName. As the number of \emph{property paths} among corresponding concepts or attributes grows, the number of mappings generated by \methodName will grow exponentially and this makes the process of selecting the best set of mappings overwhelming. However, note that in the previous experiment we have shown that \methodName is not as sensitive to the increasing number of concepts, and most large real world KBs such as medical KBs 
tend not to have a large number of properties for all concepts as we do in this case (which included an average of 24 properties for every concept in this experiment).  

In summary, we performed experiments on synthetic scenarios (inspired by existing KB exchange benchmarks~\cite{rivero2012benchmarking2}) to show how the performance of our mapping generation algorithm is affected by increasing 
complexity of the scenarios.  
Our results show that \methodName scales very well as the KB size increases, with the largest bottleneck being the number of possible interpretations of a correspondence.

\subsection{Case Study} \label{casestudy}\label{sec:casestudy}

In  
this study,
we make use of a scenario from OntoMerge~\cite{dou2005ontology}.
It includes
their  
Yale bibliography ontology as our source, and their CMU 
bibliography ontology as our target. 
We also use 
their manually curated  
mapping rules 
as our gold standard. 
We  manually identified the set of correspondences between these two ontologies (from the gold standard mapping rules).  
These ontologies are very simple, the source contains only 6 classes and no object properties, the target contains 9 classes and 3 object properties. In Phase I of our experiment, we  
asked two collaborators, \emph{c1} and \emph{c2} 
(both 
experienced 
in SPARQL) to create  mapping rules by manually writing  SPARQL queries.
We gave each collaborator 20 minutes to familiarize themselves with the ontologies and one hour to write the mapping rules.  
The queries created by \emph{c1} were able to transfer 78$\%$ of the facts  
while queries created by \emph{c2} transferred 61$\%$ of the facts. In  
Phase II, we used a ranked 
list of examples created by Kensho 
and gave  
\emph{c1} and \emph{c2} each 30 minutes to go through 
them and choose desirable examples. Note that examples were created based on query solutions obtained by running queries using query pattern of each renaming.  
Both \emph{c1} and \emph{c2} 
identified all (and only) the desirable examples.  
In  
Phase III we gave them 20 minutes to 
update  
the queries they generated in Phase I, 
based on what they  
learned from the examples. Collaborator \emph{c2}  
was able to write all the queries correctly while \emph{c1} was able to make changes  
such that 92\% of the facts translated correctly.
It is possible that \emph{c1} and \emph{c2} were able to more easily identify examples from \methodName's ranked list because in Phase I they had each tried to generate their own mapping rules. To account for this,  
we 
asked a third collaborator \emph{c3} 
(also with SPARQL experience) 
to first  
select desirable  
examples from \methodName's automatically generated list. Similar to \emph{c1} and \emph{c2}, \emph{c3} was given 30 minutes and 
was also able to pick the desirable (and only desirable) examples. We then  
gave \emph{c3} one hour to write  
queries that can transfer data from the source to the target.  
The resulting queries  
were able to transfer only 78$\%$ of the facts. 

In summary, all participants were able to correctly  
identify the desirable examples (including the  
correct optional combinations) which highlights the fact that it is more intuitive for humans to  
select from examples  
than  
to write queries manually.

\section{Related Work}\label{sec:rw}
Among mapping generation tools only
Maponto~\cite{an2006discovering}, which maps between a single relational table to a KB, traverses all paths of the target ontology to find semantic associations among concepts (as \methodName does for both source and target). 
Even to map a single table, Maponto relies heavily on the existence of enriched ontological constraints (such as cardinality) to narrow down the search over all paths.  These constraints are rarely present in real KBs.

The first step of any data sharing task is {\bf alignment}, which is the task of finding a set of suitable correspondences between the source and the target.
The second step involves interpreting a set of candidate correspondences \emph{collectively} to solve a
specific data sharing task (e.g., data exchange)~\cite{Euzenat:2013:OM:2560129}.
The output of most 
alignment tools (Step 1),  
are simple correspondences, each specifying that a resource in the source (or multiple resources like a path) has some set-theoretic relationship 
to a resource (resources) in the target. We 
take this output (from Step 1) as our input and 
find queries that collectively interpret 
correspondences. 
Ontology and schema structure have been used extensively in the first step~\cite[et al.]{Cruz2009AgreementMakerEM,Euzenat:2004:SOA:3000001.3000070,hu2005gmo,hu2008matching,li2009rimom,madhavan2001generic,melnik2002similarity,noy2001anchor,Pinkel:2013:IPY:2874493.2874497}.  
In contrast to these approaches, 
we assume the correspondences are given as {\em input} and use the semantics and structure of the KB to create (often complex) data exchange queries that can be used to correctly translate a full source instance into a full target instance without losing or changing the semantics of the data and the way values are connected. 
Note that although many heuristics for ranking correspondences are proposed~\cite{Euzenat:2013:OM:2560129}, 
they are not applicable here since they are mostly designed to express the degree of similarity of two resources and cannot rank the complex queries discovered by \methodName.
Nevertheless, the aggregation methods used to combine various integration  rankings~\cite{GalRS18,RadwanPSY09} could potentially provide insight on aggregating the rankings of the \methodName mappings.

Similar to our approach, rule mining approaches~\cite{galarraga2015fast,meng2015discovering,ortona2018robust}) traverse a knowledge graph to associate a set of resources, but they do this guided by a set of positive and/or negative examples.
We aim to find rules among corresponding resources of two different KBs, so we find associations in the source, associations in the target, and then we create our rules by combining these associations using the constraints imposed by the correspondences. On the other hand, the discovery phase of most rule mining algorithms try to fit a set of given examples. 
It would be interesting to investigate how rule mining approaches can help in enriching the set of constraints that we use in order to further refine our mapping rules. It is also interesting to see how these rules can help in ranking the mapping rules by helping to find \emph{more important} paths which fit certain sets of examples.

\section{Conclusion}\label{conclusion}
\label{sec:conclusion}

We introduced \methodName, a tool that translates knowledge between two KBs by generating mapping rules between them.
\methodName is the first KB mapping tool that is effective even when there are missing object property correspondences.
\methodName is also the first to take advantage of correspondences between object or data paths, if they are available, though our approach also allows these to be incomplete.  
\methodName improves upon existing methods by producing mappings that perform value invention in a principled way without assuming a complete source KB.

We are currently extending
\methodName in the spirit of the \emph{integration by example paradigm}~\cite{Yan:2001:DUR:375663.375729}.  
A large body of traditional data exchange literature is dedicated to identifying or exploiting examples that can shown to data engineers and incorporating their feedback~\cite{alexe2008muse,Alexe:2011:DRS:1989323.1989338,Bonifati:2008:SMV:1353343.1353358,ten2015approximation,Gottlob:2010:SMD:1667053.1667055,kimmig2018collective,Qian:2012:SSM:2213836.2213846,Yan:2001:DUR:375663.375729}.
Similarly,  
we are working on approaches for  
incorporating user feedback to improve upon our mapping rules.

When source or target instances are available,
\methodName can simplify the {\finalAssociation}s by running queries which are built using the graph pattern representation of the {\finalAssociation}.  We can  remove the parts of the query in which the variables are not being bound.
In this mode, \texttt{OPTIONAL} keywords can be removed if adding them to the query does not exchange additional facts. Note that 
all of our experiments are performed without this feature. 
Although using the above technique can help, the queries generated by our tool might still contain a considerable number of \texttt{OPTIONAL} clauses. Thus, we believe our approach can benefit from research on how to optimize the execution of SPARQL \texttt{OPTIONAL} queries~\cite{xiao2018efficient}.

As part of our ongoing work, we are exploring ideas on dealing with scale using slicing and
we plan to investigate whether
sophisticated methods such as modularization ~\cite{Ghazvinian:2011:MMU:1999676.1999684,Grau:2007:JRA:1242572.1242669,grau2008modular,konev2013model,stuckenschmidt2009modular} or partitioning~\cite{grau2005automatic,Seidenberg:2006:WOS:1135777.1135785,stuckenschmidt2004structure} may  help in dealing with knowledge translation between large KBs.
We will also consider how the process of mapping generation can be refined in the presence of other constraints such as functionality or cardinality~\cite{ppRef}. Ontology based data access initiatives (OBDA)~\cite{xiao2018ontology} facilitate the exchange and integration of data between a relational source and a target KB. It is interesting to see how our approach can be adopted in such settings.

\section{Acknowledgments}
This research was funded in part by an NSERC Strategic
Partnership Grant.\\\\

\bibliographystyle{abbrv}
\bibliography{thesis} 

\end{document}